\documentclass[11pt]{article}

\usepackage{amssymb}
\usepackage{amsmath}
\usepackage{amscd}
\usepackage{bbm}
\usepackage{latexsym}
\usepackage{ifthen}
\usepackage[xdvi]{graphics}

\catcode`\@=11
\def\@citex[#1]#2{%
\if@filesw \immediate \write \@auxout {\string \citation {#2}}\fi
\@tempcntb\m@ne \let\@h@ld\relax \def\@citea{}%
\@cite{%
  \@for \@citeb:=#2\do {%
    \@ifundefined {b@\@citeb}%
      {\@h@ld\@citea\@tempcntb\m@ne{\bf ?}%
      \@warning {Citation `\@citeb ' on page \thepage \space undefined}}%
      {\@tempcnta\@tempcntb \advance\@tempcnta\@ne%
      \@tempcntb\number\csname b@\@citeb \endcsname \relax%
      \ifnum\@tempcnta=\@tempcntb 
        \ifx\@h@ld\relax%
          \edef \@h@ld{\@citea\csname b@\@citeb\endcsname}%
        \else%
          \edef\@h@ld{\ifmmode{-}\else--\fi\csname b@\@citeb\endcsname}%
        \fi%
      \else
        \@h@ld\@citea\csname b@\@citeb \endcsname%
        \let\@h@ld\relax%
      \fi}%
    \def\@citea{,\penalty\@highpenalty\,}%
  }\@h@ld
}{#1}}
\def\@citeb#1#2{{[#1]\if@tempswa , #2\fi}}
\def\@citeu#1#2{{$^{#1}$\if@tempswa , #2\fi }}
\def\@citep#1#2{{#1\if@tempswa , #2\fi}}

\def\bcites{         
        \catcode`\@=11
        \let\@cite=\@citeb
        \catcode`\@=12
}

\def\upcites{         
        \catcode`\@=11
        \let\@cite=\@citeu
        \catcode`\@=12
}

\def\plaincites{      
        \catcode`\@=11
        \let\@cite=\@citep
        \catcode`\@=12
}
\catcode`\@=12

\catcode`@=11
\renewcommand{\section}{\@startsection{section}{1}{0pt}{\medskipamount}
{\medskipamount}{\large\bf}}
\numberwithin{equation}{section}
\catcode`@=12

\def\a{\alpha}
\def\b{\beta}
\def\g{\gamma}
\def\de{\delta}

\def\ve{\varepsilon}

\def\h{\eta}
\def\th{\theta}
\def\k{\kappa}
\def\l{\lambda}

\def\s{\sigma}

\def\vp{\varphi}

\def\ps{\psi}

\def\S{\Sigma}

\newcommand{\Om}{\Omega}
\newcommand{\C}{\mathbbm{C}}
\newcommand{\R}{\mathbbm{R}}

\newcommand{\Hcal}{{\cal H}}

\newcommand{\Ical}{{\cal I}}

\def\>{\rangle}
\def\<{\langle}
\def\N2{N=2}
\def\pa{\partial}

\def\sfrac#1#2{{\textstyle\frac{#1}{#2}}}

\newcommand{\ab}{{\bar{a}}}

\newcommand{\Zb}{\bar{Z}}

\newcommand{\Vt}{\widetilde{V}}

\newcommand{\Gt}{\widetilde{G}}

\newcommand{\ic}{\text{i}}

\newcommand{\Gp}{\ifthenelse{\boolean{mmode}}{{G^+}}{\mbox{$G^+\:$}}}
\newcommand{\Gtp}{\ifthenelse{\boolean{mmode}}{\mbox{$\Gt^+$}}{\mbox{$\Gt^+\:$}}}
\newcommand{\Gm}{\ifthenelse{\boolean{mmode}}{{G^-}}{\mbox{$G^-\:$}}}
\newcommand{\Gtm}{\ifthenelse{\boolean{mmode}}{\mbox{$\Gt^-$}}{\mbox{$\Gt^-\:$}}}

\newcommand{\uv}{\text{$|\;\!\!\!\uparrow\>$}}
\newcommand{\dv}{\text{$|\;\!\!\!\downarrow\>$}}

\newcommand{\dvb}{\text{$\<\downarrow\:\!\!\!|$}}

\newcommand{\uudvb}{\<\uparrow\uparrow\downarrow\;\!\!\!|}
\newcommand{\uduvb}{\<\uparrow\downarrow\uparrow\;\!\!\!|}
\newcommand{\duuvb}{\<\downarrow\uparrow\uparrow\;\!\!\!|}

\newcommand{\sap}{\text{\raisebox{-.5mm}{$\sqrt{{\a'}}$}}}
\newcommand{\dk}{d_\k^{\phantom{\dag}}}
\newcommand{\ddk}{d_\k^\dag}
\newcommand{\ck}{c_\k^{\phantom{\dag}}}
\newcommand{\cdk}{c_\k^\dag}
\newcommand{\reg}{R}

\begin{document}
\begin{titlepage}
\setcounter{page}{0}
\begin{flushright}
{\tt hep-th/0312314}\\
MIT--CTP--3459\\
ITP--UH--36/03\\
\end{flushright}

\vskip 2.0cm

\begin{center}

{\Large\bf String Field Theory Projectors for Fermions of Integral Weight}

\vspace{14mm}

{\large Matthias Ihl$^a$, \ Alexander Kling$^a$ \ and \ Sebastian Uhlmann$^b$}
\\[5mm]
{\em $^a$ Institut f\"ur Theoretische Physik \\
Universit\"at Hannover \\
30167 Hannover, Germany }\\[2mm]
{E-mail: {\tt msihl, kling@itp.uni-hannover.de}}\\[3mm]
{\em $^b$ Center for Theoretical Physics \\
Massachusetts Institute of Technology \\
Cambridge, MA 02139, USA}\\[2mm]
{E-mail: {\tt uhlmann@lns.mit.edu}}

\end{center}

\vspace{2cm}

\noindent
{\sc Abstract:} \\[.7cm]
The interaction vertex for a fermionic first order system of weights 
$(1,0)$ such as the twisted $bc$~system, the fermionic part of N=2 string 
field theory and the auxiliary $\eta\xi$~system of N=1 strings is 
formulated in the Moyal basis. In this basis, the Neumann matrices are 
diagonal; as usual, the eigenvectors are labeled by $\kappa\in\R$. Oscillators
constructed from these eigenvectors make up two Clifford algebras for each
nonzero value of $\kappa$.  
Using a generalization of the Moyal-Weyl map to the
fermionic case, we classify all projectors of the star-algebra which
factorize into projectors for each $\k$-subspace. 
At least for the case of squeezed states we recover 
the full set of bosonic projectors with this property. Among the subclass of
ghost number-homogeneous squeezed state projectors, we find a single
class of BPZ-real states parametrized by one (nearly) arbitrary function
of~$\k$. This class is shown to contain the generalized butterfly
states. 
Furthermore, we elaborate on sufficient and necessary
conditions which have to be fulfilled by our projectors in order to
constitute surface states.
As a byproduct we find that the full star product of N=2 string
field theory translates into a canonically normalized continuous tensor
product of Moyal-Weyl products up to an overall normalization. The
divergent factors arising from the translation
to the continuous basis cancel between bosons and fermions in any even
dimension. 

\vfill

\textwidth 6.5truein

\end{titlepage}


\section{Introduction}

\noindent
Over the last years, open string field theory has provided evidence that
it captures essential off-shell string physics, namely the process of
tachyon condensation on unstable D-brane systems (for reviews
see~\cite{Ohmori:2001am,DeSmet:2001af,Arefeva:2001ps,Taylor:2002uv,Taylor:2003gn}). This applies to cubic bosonic open string field
theory~\cite{Witten:1985cc} as well as to its supersymmetric
generalization~\cite{Witten:1986qs} and Berkovits' superstring field
theory~\cite{Berkovits:1995ab}. Most of this evidence relies heavily
on numerical work. A fully satisfactory analytic solution to the
equations of motion of open string field theory in any of its guises
is currently missing.

In order to overcome this rather unsatisfactory situation, a version of
open string field theory around the open bosonic string tachyon vacuum
has been advocated in~\cite{Rastelli:2000hv}. The key assumption in this
proposal is the pure ghost nature of the kinetic operator reflecting the
absence of physical open string excitations. In turn, this assumption
leads to a factorization of the equations of motion into matter and ghost
parts.

The matter part of these equations is a projector condition. An
enormous amount of work has been done in the meantime to identify
projectors of the star algebra in the matter 
sector~\cite{Rastelli:2001rj,Gross:2001rk,Rastelli:2001vb,Gross:2001yk,%
Moore:2001fg,Marino:2001ny,Schnabl:2002ff,Gaiotto:2002kf,Arefeva:2002sg,%
Fuchs:2002zz,Mamone:2003gu} and to
describe their properties. Apart from the identity string field, the most 
prominent representatives are the sliver and the so-called butterfly. The 
latter ones belong to a one-parameter family of surface states dubbed 
generalized butterfly states.

The ghost part of the equations of motion remain complicated to
solve (see, e.\,g., \cite{Kostelecky:2000hz,Hata:2001sq}). However,
projectors also appear in this context. In~\cite{Gaiotto:2001ji} an
auxiliary boundary conformal field theory was introduced in order
to construct solutions to the ghost equations of motion in terms of
surface state projectors of the so-called twisted $bc$ system. This
BCFT is obtained by twisting the energy momentum tensor with the
derivative of the ghost number current.\footnote{For a recent discussion
based on a regulated version of the twisted butterfly state see~%
\cite{Okawa:2003cm} and in terms of a deformed sliver state see~%
\cite{Bonora:2003cc}. For an attempt to solve Witten's cubic open string
field theory using the butterfly state see~\cite{Okawa:2003zc}.}

Eventually, we point out that projectors appear naturally in certain
solution generating techniques proposed to solve the equations of motion
of string field theory~\cite{Lechtenfeld:2000qj,Lechtenfeld:2002cu,%
Kling:2002ht,Kling:2002vi,Kawano:2001fn}.

The subject of most of the references mentioned above are surface state
projectors~\cite{LeClair:1988sp,LeClair:1988sj}. These are projectors with
field configurations arising from path integrations over fixed Riemann
surfaces whose boundary consists of a parametrized open string and a piece
with open string boundary conditions. A very fundamental result in this
realm~\cite{Gaiotto:2002kf} was that all such surfaces with the property
that their boundaries touch the midpoint of the open string lead to projector
functionals in the star algebra. Although this class of projectors is quite
large, it is advantageous to look for projectors without these singular
property which could eventually lead to D-brane solutions with finite
energy densities.

This is the motivation for the present paper. We study projectors of a
fermionic first order system of weights~$(1,0)$. It is shown that projectors
of this system give rise to bosonic projectors. Such a first order system
appears in bosonic string field theory as the twisted $bc$ system, in
the fermionic sector of N=2 string field theory~\cite{Berkovits:1997pq}
and as the auxiliary $\eta\xi$ ghost system of N=1 strings which is
introduced in the process of fermionization of the superconformal ghosts~%
\cite{Friedan:1985ge}. We use a reformulation of the interaction vertex for
such a system, which has been constructed explicitly in~\cite{Kling:2003sb}
(see also~\cite{Maccaferri:2003rz}), in terms of a continuous Moyal basis~%
\cite{Douglas:2002jm,Belov:2002fp,Arefeva:2002jj,Erler:2002nr,Belov:2002pd,%
Belov:2003df,Belov:2003qt}.\footnote{The Witten vertex was
first reformulated in terms of Moyal products in~\cite{Bars:2001ag,%
Bars:2002nu} using a discrete basis. In this paper, we find it useful to work 
in the continuous Moyal basis which is equivalent to the discrete basis.}
The diagonalization of the vertex is greatly simplified due to an intimate
relation of the bosonic matter Neumann matrices and those for the fermionic 
$(1,0)$-system~\cite{Kling:2003sb,Maccaferri:2003rz}. In the diagonal
basis, the star algebra decomposes into a product of infinitely many pairs
of Clifford algebras labeled by a parameter~$\k\in\R_+$. We employ a
fermionic version of the Moyal-Weyl map known from noncommutative field
theories to map the generators of these Clifford algebras to operators on
some auxiliary Fock space; this enables us to determine all projectors in
the star algebra which are already projectors for each value of~$\k$. Among
the general classes of projectors obtained by this procedure we find a
single class of BPZ-real squeezed state projectors which are neutral
w.\,r.\,t.\ the $U(1)$ current of the first order system. This class is
naturally parametrized by a single odd and integrable function of the
parameter $\k$ and contains the generalized butterfly states as a
subclass. It is demonstrated how to determine surface states in this
class.\footnote{Under the proviso that the surface state projector
factorizes into projectors for each value of~$\k$, they all have to reside
in this class.} A condition on the maps (conventionally denoted by~$f$)
defining the shape of the surface is given.

As a byproduct we find that the full star product of N=2 string field
theory translates into a {\it canonically normalized}\/ continuous tensor
product of Moyal-Weyl products up to an overall normalization. The
infinite factors arising from the translation to the continuous Moyal
basis cancel between bosons and fermions in any even dimension. This is
another direct consequence of the intimate relation between the bosonic and
fermionic Neumann coefficients mentioned above and is similar to an
analogous cancellation of anomalies observed in~\cite{Kling:2003sb}.

The paper is organized as follows: We start with a brief introduction
to the fermionic first order system and its connection to the bosonic
system in section~\ref{sec:moyalvertex}. This relation can be used to
immediately diagonalize the interaction vertex. In the diagonal basis,
a comparison of the integral kernel of the star product with that of
a canonical Clifford algebra product determines the (fermionic) Moyal-Weyl
pairs. It is shown that the infinite factors from the translation to
this basis cancel between bosons and fermions in N=2 string field theory
in any even dimension, in accordance with the central charge $c=-2$
of the first order system. In section~\ref{sec:projec}, we apply the
Moyal-Weyl map in the way described above and classify all projectors
which are already projectors for each value of~$\k$. The reality condition
excludes some of these classes; again only a subset consists of
squeezed state projectors. In section~\ref{sec:sqstate}, we derive a
condition on our coefficient functions such that the corresponding
states are squeezed states. It turns out that there is only one class
of real squeezed state projectors which are also neutral w.\,r.\,t.\ the
$U(1)$ ``ghost number'' current of the first order system. It is
demonstrated that the generalized butterfly states are contained in this
class. An inverse transformation from string fields to operators in the
auxiliary Fock space is given; it facilitates the computation of star
products considerably. Moreover, from squeezed state projectors of this
type one can recover the full set of bosonic projectors satisfying the 
same factorization properties. 
In section~\ref{sec:surfst}, we outline the general
method to determine whether a given squeezed state is a surface state.
We find a condition on the map~$f$ defining the surface state which
is necessary in order for it to define a surface state projector with
the above-mentioned factorization property. We then demonstrate the
procedure in the case of the generalized butterfly states. Finally, we
offer some concluding remarks in section~\ref{sec:concl}.

\section{The interaction vertex in Moyal form}\label{sec:moyalvertex}

\noindent
In this section we reformulate the interaction vertex of the fermionic
first order system~\cite{Kling:2003sb} in the continuous Moyal basis.
General methods for the diagonalization of the fermionic vertex were presented
in \cite{Bars:2003gu,Belov:2003df,Maccaferri:2003rz}. However, it seems 
advantageous 
to exploit the relation between the Neumann coefficients for the bosons and
those for the fermionic system with weights (1,0). To begin with we review
the results for the bosonic Neumann coefficients.

\noindent
{\bf Diagonalization of the bosonic coefficients.} The bosonic interaction
vertex in momentum basis is given by~\cite{Gross:1986ia,Rastelli:2001jb}
\begin{equation}
  \<V_3|=\int d^D p^{(1)}d^D p^{(2)}d^D p^{(3)}\de^D(p^{(1)}+p^{(2)}
    + p^{(3)}) {}^{(123)}\<p,0|\exp \big[-V\big]\, ,
\end{equation}
where
\begin{equation}
  V=\frac{1}{2}\sum_{r,s}\sum_{m,n\ge 1}\h_{\mu\nu}a_m^{(r)\mu}V_{mn}^{rs}
    a_n^{(s)\nu} + \sap\sum_{r,s}\sum_{n\ge 1}\h_{\mu\nu}p^{(r)\mu}
    V_{0n}^{rs}a_n^{(s)\nu}+\frac{\a'}{2}\sum_{r}\h_{\mu\nu}p^{(r)\mu}
    V_{00}^{rr}p^{(r)\nu}\, .
\end{equation}
Here $a_m^\mu$ are the bosonic oscillators normalized such that $[a_m^\mu,
a_n^\nu]=\de_{m,-n}\eta^{\mu\nu}$, and $p^\mu$ is the center of mass
momentum. The Neumann coefficients $V_{mn}^{rs}$ were given in terms of
coefficients of generating functions in~\cite{Gross:1986ia,%
Rastelli:2001jb}. They were diagonalized in~\cite{Rastelli:2001hh}; the
common eigenvectors $v_n(\k)$ for $M^{rs}_{mn}:=CV^{rs}_{mn}$ with $1\leq
r,s\leq 3$, $C_{mn}=(-1)^m\de_{m,n}$ are labeled by a continuous parameter
$\k\in\R$, i.\,e.,
\begin{equation}
  M^{rs}_{mn}v_{n}(\k)=\mu^{rs}(\k)\, v_m(\k)\, .
\end{equation}
Here and in the following, a summation from $1$ to $\infty$ over repeated
indices is implied. The eigenvalues were found to be
\begin{subequations}\label{eq:rszevs}
\begin{align}
  \mu^{11}(\k) &= - \frac{1}{1+2 \cosh\frac{\pi \k}{2}}\, , \\
  \mu^{12}(\k) &= \frac{1 + \cosh\frac{\pi \k}{2} + \sinh\frac{\pi \k}{2}}
    {1+2 \cosh\frac{\pi \k}{2}}\, , \\
  \mu^{21}(\k) &= \frac{1 + \cosh\frac{\pi \k}{2} - \sinh\frac{\pi \k}{2}}
    {1+2 \cosh\frac{\pi \k}{2}}\, .
\end{align}
\end{subequations}

\noindent
{\bf Diagonalization of the fermionic coefficients.} It was shown that
solutions to the ghost part of the vacuum string field theory equations
may be obtained from certain projectors of the twisted $bc$ system~%
\cite{Gaiotto:2001ji}. In this theory, the energy-momentum tensor of the
untwisted $bc$ system for the critical bosonic string is supplemented
by $-\pa J_{bc}$, the negative derivative of the ghost number current.
This shifts the weights of $b$ to~$1$ and of~$c$ to~$0$, resulting in a
first order system $b'c'$ of central charge $c_{b'c'}=-2$.

It turns out~\cite{Kling:2003sb} that this first order system is
exactly the fermionic part of string field theory for strings with
N=2 world-sheet supersymmetry. This theory possesses a twisted N=4
superconformal invariance on the world-sheet and describes the dynamics
of N=2 strings in a (four-dimensional) K\"ahler space-time; therefore,
two complex bosons $Z^a$ (with $a=1,2$) and their twisted
superpartners $\ps^{+a}$ and $\ps^{-\ab}$ are needed. For each $a$,
it was shown in~\cite{Kling:2003sb} that the fermionic $\ps^{+a}
\ps^{-\ab}$ system coincides with the twisted $bc$ system described
above.\footnote{As indicated by the central charges, the reparametrization
anomalies of one $\ps^\pm$ system and one $Z\Zb$ system exactly cancel.}
For the following, it is thus irrelevant whether we are speaking of
twisted $bc$ systems or of $\ps^\pm$ systems. For convenience, we mostly
choose the latter alternative. As a trivial difference, however, it should
be noted that we normalize the anticommutation relations for the $\ps^\pm$
system in such a way that they match complex fermions,
\begin{equation}
  \{\ps^{+a}_m, \ps^{-\ab}_n\} = 2\eta^{a\ab}\de_{m,-n} \label{eq:comrel}
\end{equation}
for a K\"ahler metric with components $\eta^{a\ab}$, whereas the
normalization for the twisted $bc$ system conventionally differs from
this by a factor of~2.

As fields of integral weight, both $\ps^+$ and $\ps^-$ are integer-moded.
In particular, the spin~0 field $\ps^+$ has a zero-mode on the sphere. In
analogy to the $bc$ system there are thus two vacua at the same energy
level: the bosonic $SL(2,\R)$-invariant vacuum $|0\>=:\dv$ is annihilated
by the Virasoro modes $L_{m\geq -1}$ and $\ps^+_{m>0}$, $\ps^-_{m\geq 0}$;
its fermionic partner, $\uv:=\ps^+_0\dv$, is annihilated by $\ps^+_{m\geq
0}$, $\ps^-_{m>0}$. To get nonvanishing fermionic correlation functions, we
need one $\ps^+$-insertion, i.e., $\<\downarrow\!\dv = \<\uparrow\!\uv = 0$,
$\<\downarrow\!\uv = 1$.

The interaction vertex for the fermionic $\ps^\pm$ system is then given by
\begin{equation}
  \<V_3| = \big( \uudvb+\uduvb+\duuvb \big) \exp\big[ \sfrac{1}{4}\sum_{r,s}
    \sum_{k=1,l=0}^\infty \ps^{+(r)}_k N_{kl}^{rs} \ps^{-(s)}_l\big]
    \label{eq:psivrtx}
\end{equation}
in terms of Neumann coefficients $N_{kl}^{rs}$ with $1\leq r,s\leq 3$. For
the nonzero-mode part of these coefficients it was shown in~%
\cite{Kling:2003sb} (cf.\ also~\cite{Maccaferri:2003rz}) that they are
related to the bosonic coefficients $N^{rs}_{mn}$ by
\begin{equation}
  N^{rs}_{mn} = 2\sqrt{\frac{m}{n}} V^{rs}_{mn}\, , \quad m,n\ge 1\, .
\label{eq:bosfer}
\end{equation}
This renders the diagonalization almost trivial. Consider a matrix $U$
diagonalizing $V^{rs}_{mn}$ with $m,n\geq 1$ into a matrix $D$ by adjoint
action. Then one can immediately diagonalize the nonzero-mode part of
$N^{rs}_{mn}$ as
\begin{equation}
  D_{kl} = U^{-1}_{km}2\;\! V^{rs}_{mn}U_{nl} = U^{-1}_{km}
    \Big( \frac{1}{\sqrt{E}}N^{rs}\sqrt{E}\Big)_{mn} U_{nl} =:
    {U'}^{-1}_{km}N^{rs}_{mn}U_{nl}'\, ,
\label{eq:diagU}
\end{equation}
where we have introduced the matrix $E_{mn}=m\,\de_{mn}$. Obviously,
$V^{rs}_{mn}$ and $N^{rs}_{mn}$ share the same eigenvalues~%
(\ref{eq:rszevs}) (up to a factor of~2),\footnote{Since $N_{0n}^{rs}=0$,
it is obvious that the inclusion of the zero-mode part (i.\,e.,
admitting $m,n\geq 0$) simply adds an eigenvalue~$0$ to the spectrum.}
and it follows that
\begin{subequations}
\begin{align}
  \Big(\sqrt{E}\, 2 M^{rs}\frac{1}{\sqrt{E}}\Big)_{mn}\Big(\sqrt{E}\cdot
    v(\k)\Big)_n & = 2\mu^{rs}(\k)\,\Big(\sqrt{E}\cdot v(\k)\Big)_m\, , \\
  \Big(v(\k)\cdot\frac{1}{\sqrt{E}}\Big)_m\Big(\sqrt{E}\, 2 M^{rs}\frac{1}
    {\sqrt{E}}\Big)_{mn} & = 2\mu^{rs}(\k)\,\Big(v(\k)\cdot\frac{1}
    {\sqrt{E}}\Big)_n\, ,
\end{align}
\end{subequations}
which suggests the definitions
\begin{equation}
  v_m^+(\k) := \Big(\sqrt{E}\cdot v(\k)\Big)_m = \sqrt{m}\, v_m(\k)\, ,
    \quad
  v_m^-(\k) := \Big(\frac{1}{\sqrt{E}}\cdot v(\k)\Big)_m =
    \frac{1}{\sqrt{m}}\,v_m(\k)\, . \label{eq:v+-}
\end{equation}
From~\cite{Douglas:2002jm} one derives the obvious relations
\begin{equation}\label{eq:fEVrel}
  \int_{-\infty}^{\infty}d\k\,v_m^+(\k)v_n^-(\k) = \de_{mn}\, , \quad
    \sum_m\,v_m^+(\k)v_m^-(\k') = \de(\k-\k')\, ,
\end{equation}
which can be split into even and odd parts to give
\begin{subequations}
\label{eq:orthcmpl}
\begin{align}
  2\int_0^\infty d\k\,v_{2m}^+(\k)v_{2n}^-(\k) & = \de_{mn}\, , &
  2\int_0^\infty d\k\,v_{2m+1}^+(\k)v_{2n+1}^-(\k) & = \de_{mn}\, , \\
  2\sum_{n=1}^\infty v_{2n}^+(\k) v_{2n}^-(\k') & = \de(\k-\k')\, , &
  2\sum_{n=1}^\infty v_{2n-1}^+(\k) v_{2n-1}^-(\k') & = \de(\k-\k')
\end{align}
\end{subequations}
for $\k>0$. They are given by the generating functionals
\begin{subequations}
\label{eq:eogenf}
\begin{align}
  f_{v_e^-}(\k,z) & = \sum_{n=1}^\infty v_{2n}^-(\k) z^{2n} = \frac{1}
    {\sqrt{{\cal N}(\k)}\k}(1-\cosh\k Z)\, , \\
  f_{v_o^-}(\k,z) & = \sum_{n=1}^\infty v_{2n-1}^-(\k) z^{2n-1} =
    \frac{1}{\sqrt{{\cal N}(\k)}\k}\sinh\k Z\, , \\
  f_{v_e^+}(\k,z) & = \sum_{n=1}^\infty v_{2n}^+(\k) z^{2n} = -\frac{1}
    {\sqrt{{\cal N}(\k)}}\frac{z}{1+z^2}\sinh\k Z\, , \\
  f_{v_o^+}(\k,z) & = \sum_{n=1}^\infty v_{2n-1}^-(\k) z^{2n-1} =
    \frac{1}{\sqrt{{\cal N}(\k)}}\frac{z}{1+z^2}\cosh\k Z
\end{align}
\end{subequations}
with $Z:=\tan^{-1} z$ and ${\cal N}(\k)=\frac{2}{\k}\sinh\frac{\pi\k}{2}$.
We define the continuous modes
\begin{subequations}\label{eq:contbas}
\begin{align}
  \psi_{e,\k}^{-\dag} & := \sqrt{2}\sum_{n=1}^{\infty}v_{2n}^-(\k)
    \psi_{-2n}^-\, , &
  \psi_{o,\k}^{-\dag} & := -\sqrt{2}\ic\sum_{n=1}^{\infty}
    v_{2n-1}^-(\k)\psi_{-2n+1}^-\, , \\
  \psi_{e,\k}^{+\dag} & := \sqrt{2}\sum_{n=1}^{\infty}v_{2n}^+(\k)
    \psi_{-2n}^+\, , &
  \psi_{o,\k}^{+\dag} & := -\sqrt{2}\ic\sum_{n=1}^{\infty}
    v_{2n-1}^+(\k)\psi_{-2n+1}^+\, .
\end{align}
\end{subequations}
The factors of $\ic$ are chosen in such a way that the BPZ conjugate of
$\psi_{o,\k}^+$ $(\psi_{o,\k}^-)$ coincides with (minus) the hermitean
conjugate; similar relations are true for the even components.
They satisfy the anticommutation relations
\begin{equation}
  \{\psi_{e,\k}^{+\dag},\psi_{e,\k'}^-\}=2\,\de(\k-\k')\, ,\qquad
    \{\psi_{o,\k}^{+\dag},\psi_{o,\k'}^-\}=2\,\de(\k-\k')
\end{equation}
following from the anticommutation relations of $\psi_n^+$ and
$\psi_n^-$ and the completeness relations of the eigenvectors
$v_n^+(\k)$ and $v_n^-(\k)$. The relations~(\ref{eq:contbas})
can be inverted:
\begin{subequations}\label{eq:invcontbas}
\begin{align}
  \psi_{-2n}^- & = \sqrt{2}\int_0^{\infty}d\k\; v_{2n}^+(\k)
    \psi_{e,\k}^{-\dag}\, , &
  \psi_{-2n+1}^- & = \sqrt{2}\ic \int_0^{\infty}d\k \; v_{2n-1}^+(\k)
    \psi_{o,\k}^{-\dag}\, , \\
  \psi_{-2n}^+ & = \sqrt{2} \int_0^{\infty}d\k \; v_{2n}^-(\k)
    \psi_{e,\k}^{+\dag}\, , &
  \psi_{-2n+1}^+ & = \sqrt{2}\ic \int_0^{\infty}d\k \;
    v_{2n-1}^-(\k)\psi_{o,\k}^{+\dag}\, .
\end{align}
\end{subequations}
We will use the continuous modes to rewrite the interaction vertex~%
(\ref{eq:psivrtx}) in Moyal form.

\noindent
{\bf The interaction vertex in the continuous basis.} In this paper,
we restrict to the nonzero-mode part of the vertex. Only for these
modes the connection to the bosonic vertex holds, and we find a
correspondence of the projectors in both sectors. This is tantamount
to the choice of a generalization of the Siegel gauge condition
$\ps_0^-|\phi\>=0$. All known surface state projectors (which
factorize into projectors for each value of~$\k$) satisfy this
condition; they are based on the $\dv$-vacuum.

The nonzero-mode part of the ket-interaction vertex reads
\begin{equation}
  |V_3\>' = \exp\Big[-\sfrac{1}{4}\sum_{r,s}\sum_{k,l\ge1}
    \psi_{-k}^{+(r)}\,(CNC)_{kl}^{rs}\,\psi_{-l}^{-(s)}\Big]|\Om\>\, .
\end{equation}
Using the relations of the previous paragraph we can rewrite it as
\begin{equation}
\begin{split}
  |V_3\>' & =\exp\Big[-\sfrac{1}{2}\sum_{r,s}\sum_{k,l\ge1}
    \psi_{-k}^{+(r)}\,\Big(C\sqrt{E}V^{rs}\frac{1}{\sqrt{E}}C\Big)_{kl}\,
    \psi_{-l}^{-(s)}\Big]|\Om\> \\
  & =\exp\Big[-\sfrac{1}{2}\sum_{r,s}\sum_{k,l\ge1}
    \psi_{-k}^{+(r)}\,\Big(\sqrt{E}M^{rs}C\frac{1}{\sqrt{E}}\Big)_{kl}\,
    \psi_{-l}^{-(s)}\Big]|\Om\>\, .
\end{split}
\end{equation}
Splitting in even and odd parts and inserting the continuous
basis~(\ref{eq:contbas}) we obtain
\begin{multline}
  \sum_{r,s}\sum_{k,l\ge 1}\psi_{-k}^{+(r)}\,\Big(\sqrt{E}M^{rs}C\frac{1}
    {\sqrt{E}}\Big)_{kl}\,\psi_{-l}^{-(s)} \\
  = \sum_{r,s}\sfrac{1}{2}\int_0^{\infty}d\k\,\Big[ \big(\mu^{rs}(\k)+
    \mu^{sr}(\k)\big) \Big({\psi_{e,\k}^{+(r)}}^\dag
    {\psi_{e,\k}^{-(s)}}^\dag+{\psi_{o,\k}^{+(r)}}^\dag
    {\psi_{o,\k}^{-(s)}}^\dag\Big) \\
  -\big(\mu^{rs}(\k)-\mu^{sr}(\k)\big)\Big(\ic{\psi_{e,\k}^{+(r)}}^\dag
    {\psi_{o,\k}^{-(s)}}^\dag-\ic{\psi_{o,\k}^{+(r)}}^\dag
    {\psi_{e,\k}^{-(s)}}^\dag\Big)\Big] \, .
\end{multline}
For notational ease we introduce
\begin{equation}
  {\vec\psi}_\k^\pm{}^\dag = \begin{pmatrix}
       {\psi^{\pm}_{e,\k}}^\dag \\ {\psi^{\pm}_{o,\k}}^\dag
     \end{pmatrix}\, ;
\end{equation}
in this vector notation, the nonzero-mode part of the interaction
vertex eventually takes the form
\begin{equation}\label{eq:nonzero}
  |V_3\>' = \exp\Big[-\sfrac{1}{2}\sum_{r,s}\int_0^{\infty}d\k\,\Big(
    \mu^{(rs)}(\k){{\vec\psi}_{\k}^{+}}{}^{\dag(r)}\cdot
            {{\vec\psi}_{\k}^-}{}^{\dag(s)}+
    \mu^{[rs]}(\k){{\vec\psi}_{\k}^+}{}^{\dag(r)}\cdot\s_y\cdot
            {{\vec\psi}_{\k}^-}{}^{\dag(s)}\Big)\Big]|\Om\>\, ,
\end{equation}
where $\s_y$ denotes the second Pauli matrix, and $\mu^{(rs)}(\k)$ and
$\mu^{[rs]}(\k)$ are the symmetric and antisymmetric parts of
$\mu^{rs}(\k)$, respectively.

\noindent
{\bf Identification of Moyal structures.} We can now readily transform the
nonzero-mode part of the interaction vertex~(\ref{eq:nonzero}) into a
form which resembles the structure of the oscillator vertex for the
Moyal-Weyl product~\cite{Douglas:2002jm,Erler:2002nr} for anticommuting
quantities. As a first step, observe that~(\ref{eq:nonzero}) can be
rewritten as
\begin{equation}
  |V_3\>' = \exp\Big[-\sfrac{1}{2}\sum_{r,s}\int_0^{\infty}d\k\,
    {{\vec\psi}_\k^+}{}^{(r)\dag}\cdot \Vt^{rs}\cdot {{\vec\psi}_\k^-}
    {}^{(s)\dag}\Big]|\Om\>\, ,
\end{equation}
where $\Vt^{rs}$ is the $6\times6$ matrix
\begin{equation}
  \Vt^{rs}= \frac{1}{\th^2(\k)+12} \begin{pmatrix}
    \th^2(\k) - 4 & 8+ 4\,\th(\k)\s_y & 8- 4\,\th(\k) \s_y \\
    8- 4\,\th(\k) \s_y & \th^2(\k) - 4 & 8+ 4\,\th(\k) \s_y \\
    8+ 4\,\th(\k) \s_y & 8-4\,\th(\k) \s_y & \th^2(\k) -4 \end{pmatrix}^{rs},
\end{equation}
and $\th(\k)= 2\tanh\sfrac{\pi\k}{4}$ is the unique solution to the
equations
\begin{subequations}
\begin{align}
  \mu^{11}(\k) & = \frac{\th^2(\k)-4}{\th^2(\k)+12}\, , \\
  \frac{1}{2}(\mu^{12}(\k)+\mu^{21}(\k)) & = \frac{8}{\th^2(\k)+12}\, , \\
  \frac{1}{2}(\mu^{12}(\k)-\mu^{21}(\k)) & = \frac{4\,\th(\k)}
    {\th^2(\k)+12}\, .
\end{align}
\end{subequations}
As a second step, we contract the interaction vertex with eigenvectors
of the position operators $\vec{x}_{\k} = \left(
\begin{smallmatrix} x_{e,\k} \\ x_{o,\k} \end{smallmatrix}\right)
= \frac{\ic}{\sqrt{2}}(\vec{\psi}_{\k}^- -\vec{\psi}_{\k}^-{}^\dag)$
and $\vec{y}_{\k} = \left(\begin{smallmatrix} y_{e,\k} \\ y_{o,\k}
\end{smallmatrix}\right) = \frac{1}{\sqrt{2}}(\vec{\psi}_{\k}^+
+\vec{\psi}_{\k}^+{}^\dag)$ in order to obtain an integral kernel
representation for the Moyal product~\cite{Erler:2002nr}. The
eigenvectors are given by
\begin{align}
\label{eq:Xev}
\begin{split}
  \< \vec{X}| := \< \vec{x},\vec{y}|
    = \dvb\exp\Big[ -\frac{1}{2} \int_0^{\infty} d\k\, \big(
    & \vec{\psi}_\k^+\cdot \vec{\psi}_\k^-
    + \ic\sqrt{2}\vec{\psi}_\k^+{}\cdot \vec{x}(\k) \\
  & -\sqrt{2}\vec{y}(\k)\cdot \vec{\psi}_\k^-{}
    + \ic\vec{x}(\k)\cdot \vec{y}(\k) \big)\Big]\, .
\end{split}
\end{align}
Then the integral kernel for the star product is given by
\begin{align}\label{eq:kernel}
  K(\vec{X}^{(1)},&\vec{X}^{(2)};\vec{X}^{(3)}) =
    \<\vec{X}^{(3)}|\<\vec{X}^{(1)}|\<\vec{X}^{(2)}|V_3\> \\
  & = \exp\big[\de(0)\int\!d\k\,\ln\big(\det{(1+\Vt^{rs})}\big)\big]
    \exp\Big[-\ic\sum_{r,s} \int_0^{\infty} d\k\,
    \vec{x}^{(r)}(\k) \cdot W^{rs} \cdot \vec{y}^{(s)}(\k)\Big]\, ,
    \notag
\end{align}
where $W^{rs}:=\frac{1-\Vt^{rs}}{1+\Vt^{rs}}$ takes the form
\begin{equation}
  W^{rs} = \th^{-1}\begin{pmatrix} 0 & -\s_y & \s_y \\ \s_y &
    0 & -\s_y \\
    -\s_y & \s_y & 0 \end{pmatrix}^{rs}\, ,
\end{equation}
and $\det{(1+\Vt^{rs})}=\frac{64\,\th^4(\k)}{(\th^2(\k)+12)^2}$. This is
a product over Moyal kernels for a tensor product of two Clifford
algebras~\cite{Erler:2002nr} for each $\k>0$ since
\begin{equation}\label{eq:mkernel}
  -\ic \vec{x}^{(r)}(\k) W^{rs} \vec{y}^{(s)}(\k) =
    x_e^{(r)}(\k) K^{rs} y_o^{(s)}(\k) -x_o^{(r)}(\k)K^{rs}
    y_e^{(s)}(\k)\, ,
\end{equation}
where
\begin{equation}
  \label{eq:K}
  K^{rs}= \th^{-1}\begin{pmatrix} 0 & 1 & -1 \\ -1 & 0 & 1 \\ 1 & -1 & 0
          \end{pmatrix}\, .
\end{equation}
From this, we read off that for a canonically normalized star product
we can identify two separate continuous Moyal-Weyl pairs, namely
$(x_e(\k),y_o(\k))$ and $(x_o(\k),y_e(\k))$ with
\begin{equation}\label{eq:stcomm}
\begin{split}
  \{x_e(\k)\stackrel{\star}{,}y_o(\k')\} & = 2\,\th(\k)\de(\k-\k')\, ,\\
  \{x_o(\k)\stackrel{\star}{,}y_e(\k')\} & = -2\,\th(\k)\de(\k-\k')\, ,
\end{split}
\end{equation}
which is twice the noncommutativity for one real boson. All other
anticommutators vanish.\footnote{For $\k=0$, the algebra is
anticommutative.} This result can be checked if one uses the path
integral measure for which the norm of the ground state
\begin{equation}
  \label{eq:grstate}
  \Psi_{|0\>}=\exp\Big[ -\frac{\ic}{2} \int_0^{\infty} d\k\, \big(
             \vec{x}(\k)\cdot \vec{y}(\k) \big)\Big]
\end{equation}
(which is recovered from contraction of~(\ref{eq:Xev}) with the vacuum) is
one.

\noindent
{\bf Cancellation of determinants.} It is worthwhile to pause here
for a moment and have a closer look at the meaning of the result in
eq.~(\ref{eq:kernel}) for N=2 string field theory. As could have been expected
the Moyal kernel differs from the result obtained in~\cite{Erler:2002nr}
essentially by the normalization factor. For $D/2$ $\psi^{\pm}$-pairs,
i.\,e., in a spacetime of real even dimension $D$, the normalization
factor can be read off from~(\ref{eq:kernel}). The analog of the
constant $C'$ of~\cite{Douglas:2002jm} evaluates to
\begin{equation}\label{eq:detpsi}
  {\cal N}_{\psi}=\exp\big[-\de(0)\frac{D}{2}\int\!d\k\,
    \ln\big(\frac{(\th^2(\k)+12)^2}{64}\big)\big]\, .
\end{equation}
In~\cite{Douglas:2002jm} a cancellation of such normalization factors
between the matter and ghost sector was advocated. It turned
out~\cite{Erler:2002nr}
that this cancellation does not occur in bosonic string field theory,
at least for the reduced star product of the ghost sector. Here the
situation is quite different. The normalization factor for the bosonic
(matter) part of the star product computed in~\cite{Douglas:2002jm} is
proportional to the number of spacetime dimensions $D$ and reads
\begin{equation}
  {\cal N}_{X}=\exp\big[\de(0)D\int\!d\k\,
    \ln\big(\frac{1}{8}(\th^2(\k)+12)\big)\big]\, .
\end{equation}
Hence, taking into account that due to eqs. (\ref{eq:v+-}) and 
(\ref{eq:fEVrel}) the eigenvalue densities\footnote{For a 
discussion of the spectral density see~\cite{Fuchs:2002wk,Belov:2002sq}.}
for bosons and fermions are equal
\begin{equation}
  \label{eq:EVdensity}
  \rho(\k)_{X}=\sum_{n=1}^{\infty}v_n(\k) v_n(\k) =
  \sum_{n=1}^{\infty}v^-_n(\k) v^+_n(\k)=\rho(\k)_{\psi}\, ,
\end{equation}
the advocated cancellation
\begin{equation}
  {\cal{N}}_{X}{\cal{N}}_{\psi}=1
\end{equation}
indeed happens in any even dimension.\footnote{Note that this will also hold 
for the finite part of the spectral density after regularization.} This is a 
distinguished feature of N=2 string field theory. A similar cancellation has 
been observed for the anomaly of midpoint preserving reparametrizations
in~\cite{Kling:2003sb}. Both cancellations are consequences of
relation~(\ref{eq:bosfer}).

The same is true for the canonical normalizations, which are given by
\begin{equation}
  {\cal N}_{\psi,\text{can.}}=\exp\big[\de(0)\frac{D}{2}\int\!d\k\,
    \ln\big(\th^4\big)\big]
\end{equation}
for the fermions and by
\begin{equation}
  {\cal N}_{X,\text{can.}}=\exp\big[-\de(0)D\int\!d\k\,
    \ln\big(\th^2)\big]
\end{equation}
for the bosons, where the path integral measure is normalized such that
the norm of the bosonic as well as the norm of the fermionic ground state
is 1. The full Witten star product of N=2 string field theory is a
{\it canonically normalized continuous tensor product of Moyal-Weyl
products}.

\section{Star algebra projectors} \label{sec:projec}
\noindent
As known from noncommutative field theories, the computation of star
products can be facilitated considerably by making use of the Moyal-%
Weyl map. In our case, we can do the same; now, string fields are
translated into operators in some auxiliary Fock space~$\Hcal_\text{aux}$,
and star products are computed via pairs of (Grassmann-odd) pairs of
creation and annihilation operators for each~$\k$. This makes it possible
to identify a large subclass of projectors of the twisted ghost part of
the star algebra, namely those which are projectors for each~$\k>0$
separately.

\noindent
{\bf Operator representation of the star product.} The anticommutation
relations~(\ref{eq:stcomm}) suggest that the Moyal-Weyl map should have
the following properties:
\begin{align}\label{eq:MWop}
\begin{split}
  \frac{x_e(\k)}{\sqrt{2\th(\k)}}\mapsto \ck\, ,\quad
    \frac{y_o(\k)}{\sqrt{2\th(\k)}}\mapsto \cdk\, , \\
  \frac{\ic\,x_o(\k)}{\sqrt{2\th(\k)}}\mapsto \dk\, ,\quad
    \frac{\ic\,y_e(\k)}{\sqrt{2\th(\k)}}\mapsto \ddk\, ,
\end{split}
\end{align}
where the operators act in an auxiliary Fock space with vacua
\begin{equation}
  \ck|0\>_c = 0\, , \quad \dk|0\>_d = 0\, .
\end{equation}
In the following we will investigate the star product for a single~$\k$.
The Moyal-Weyl map assigns to each functional $\Phi\big(x_e(\k),x_o(\k),
y_e(\k),y_o(\k)\big)$ an operator $\hat\Phi(\ck,\dk,\ddk,\cdk)$
acting in the auxiliary Fock space. In principle, this assignment
involves a particular ordering prescription for the operators. We decide
to keep the ordering which is inherited from the expressions as string 
fields. It is understood that under the Moyal-Weyl map,
\begin{equation}
  \Phi\star\Psi\mapsto \hat\Phi\hat\Psi\, .
\end{equation}
In this way, the star anticommutators~(\ref{eq:stcomm}) are mapped to
anticommutators of operators. In order to be able to obtain reasonable
results in the continuum normalization, we have to regularize
the delta distributions in~(\ref{eq:stcomm}). To this aim, we
introduce a level regulator $L$ as in~\cite{Rastelli:2001hh,%
Okuyama:2002yr,Okuyama:2002tw} and formally set 
$\de(0)\sim\frac{\log L}{2\pi}=:\reg$. Under the Moyal-Weyl map, the 
regularized star anticommutators can be rewritten as
\begin{equation}\label{eq:MWcr}
  \{\ck,\cdk\} = \reg\, ,\qquad \{\dk,\ddk\} = \reg\, ,
\end{equation}
with all other anticommutators vanishing.

\noindent
{\bf Projectors from the Moyal-Weyl map.} We will now take advantage
of the map defined above to classify projectors for the combined system
of fermionic operators $\ck$, $\cdk$ and $\dk$, $\ddk$. It will turn out
that the combined system allows for a much larger variety of projectors as
compared to a single set of operators. Namely, for a bosonic rank one
projector built out of, say, $\ck$ and $\cdk$, one finds that the only
possibilities are  $\mathbbm{1}$, $\frac{1}{\reg}\ck\cdk$ and $\frac{1}
{\reg}\cdk\ck$. These operators correspond to $|0\>_c\;\!{}_c\<0|+|1\>_c
\;\!{}_c\<1|$, $|0\>_c\;\!{}_c\<0|$ and $|1\>_c\;\!{}_c\<1|$,
respectively, where $|1\>_c=\cdk|0\>_c$. Of course, the combined system
of oscillators will contain more projectors than the naive ones build as
products of projectors in each $\k$-subsystem; but, as mentioned above,
this paper is devoted to an investigation of states which are projectors
for each value of~$\k$.

We insert the ansatz
\begin{equation}
  \hat P_\k = \a\mathbbm{1} + \b\,\ck\cdk + \g\,\dk\ddk + \de\,\ck\dk
    + \ve\,\cdk\ddk + \vp\,\ck\ddk + \h\,\cdk\dk + \l\,\ck\cdk\dk\ddk
  \label{eq:projans}
\end{equation}
into the projector condition $\hat{P_\k}\hat{P_\k} = \hat{P_\k}$ and
read off equations for the coefficient functions. This yields the
following set of equations:
\begin{subequations}\label{eq:pc}
\begin{align}
  \a &= \a^2-\reg^2\ve\de\, , \label{eq:pc1}\\
  \b &= 2\a\b+\reg\b^2+\reg\ve\de-\reg\vp\h\, , \label{eq:pc2}\\
  \g &= 2\a\g+\reg\g^2+\reg\ve\de-\reg\vp\h\, , \label{eq:pc3}\\
 \de &= (2\a+\reg\b+\reg\g+\reg^2\l)\de\, , \label{eq:pc4}\\
 \ve &= (2\a+\reg\b+\reg\g+\reg^2\l)\ve\, , \label{eq:pc5}\\
 \vp &= (2\a+\reg\b+\reg\g)\vp\, , \label{eq:pc6}\\
  \h &= (2\a+\reg\b+\reg\g)\h\, , \label{eq:pc7}\\
  \l &= \reg^2\l^2+2(\a+\reg\b+\reg\g)\l+2\b\g-2\ve\de+2\vp\h\, .\label{eq:pc8}
\end{align}
\end{subequations}
Subtracting eq.~(\ref{eq:pc3}) from eq.~(\ref{eq:pc2}) one finds
\begin{equation}
  (\b-\g)(2\a-1+\reg\b+\reg\g)=0\, .
\end{equation}
We will use this condition to classify the solutions to~(\ref{eq:pc}):
\begin{itemize}
  \item[(I):] $2\a+\reg\b+\reg\g =1$
  \begin{itemize}
    \item[(a):] $\de\ne 0$ $\vee$ $\ve\ne 0$ $\Rightarrow$ $\l=0\,$,
                   $\de\ve=\b\g+\vp\h=\sfrac{1}{\reg^2}(\a^2-\a)\,$.
    \item[(b):] $\de=\ve=0$ $\Rightarrow$ $\a\in\{0,1\}\,$.
    \begin{itemize}
      \item[(b1):] $\a=0$ $\Rightarrow$ $\b=\sfrac{1}{\reg}-\g\,$,
        $\g(\g-\sfrac{1}{\reg})=\vp\h\,$, $\l\in\{0,-\sfrac{1}
        {\reg^2}\}\,$.
      \item[(b2):] $\a=1$ $\Rightarrow$ $\b=-\sfrac{1}{\reg}-\g\,$,
        $\g(\g+\sfrac{1}{\reg})=\vp\h\,$, $\l\in\{0,\sfrac{1}
        {\reg^2}\}\,$.
    \end{itemize}
  \end{itemize}
  \item[(II):] $\b=\g$
  \begin{itemize}
    \item[(a):] $\vp\ne 0$ $\vee$ $\h\ne 0$ $\Rightarrow$
                $\a=\sfrac{1}{2}-\reg\g\,$, $\de\ve=\g^2-\sfrac{1}
                {4\reg^2}\,$, $\vp\h=-\sfrac{1}{4\reg^2}\,$.
    \begin{itemize}
      \item[(a1):] $\de\ve\ne 0$ $\Rightarrow$ $\l=0\,$.
      \item[(a2):] $(\de\ne 0\, ,\,\ve=0)$ $\vee$ $(\de= 0\, ,\,\ve\ne 0)$
                   $\Rightarrow$ $\l=0\,$, 
                   $(\g=\sfrac{1}{2\reg}\, ,\,\a=0)$ $\vee$
                   $(\g=-\sfrac{1}{2\reg}\, ,\,\a=1)\,$.
      \item[(a3):] $\de=\ve=0$ $\Rightarrow$
        $(\g=\sfrac{1}{2\reg}\, ,\,\a=0,\,\l\in\{0,-\sfrac{1}{\reg^2}\})$
        $\vee$ $(\g=-\sfrac{1}{2\reg}\, ,\,\a=1,\, \l\in\{0,\sfrac{1}
        {\reg^2}\})\,$.
    \end{itemize}
    \item[(b):] $\vp=\h=0$ $\Rightarrow$ $\g=\sfrac{1-\a}{\reg}$ $\vee$
        $\g=-\sfrac{\a}{\reg}\,$, $\l=\sfrac{1-2\a}{\reg^2}-\sfrac{2\g}
          {\reg}$ $\vee$ $\l=-\sfrac{2\g}{\reg}\,$, $\de\ve=
          \sfrac{\a^2-\a}{\reg^2}\,$.
    \begin{itemize}
      \item[(b1):] $\de\ve\ne 0\,$.
      \item[(b2):] $(\de\ne 0\, ,\,\ve=0)$ $\vee$ $(\de= 0\, ,\,\ve\ne 0)$
           $\Rightarrow$ \\
           $(\a,\g,\l)\in\{(0,\sfrac{1}{\reg},-\sfrac{1}{\reg^2}),
            (0,0,\sfrac{1}{\reg^2}),(1,0,-\sfrac{1}{\reg^2}),
            (1,-\sfrac{1}{\reg},\sfrac{1}{\reg^2})\}\,$.
      \item[(b3):] $\de=\ve=0\:\Rightarrow\:(\a,\g,\l)\in\{(0,
        \sfrac{1}{\reg},-\sfrac{1}{\reg^2}),(0,\sfrac{1}{\reg},
        -\sfrac{2}{\reg^2}),(0,0,\frac{1}{\reg^2}),(0,0,0),\\
        \phantom{\de=\ve=0\:\Rightarrow\:(\a,\g,\l)\in\{}(1,0,
        -\sfrac{1}{\reg^2}),(1,0,0),(1,-\sfrac{1}{\reg},\sfrac{1}
        {\reg^2}),(1,-\sfrac{1}{\reg},\sfrac{2}{\reg^2})\}\,$.
    \end{itemize}
  \end{itemize}
\end{itemize}
The first and second level (denoted by roman numbers and latin letters)
specify general conditions which have to be fulfilled
simultaneously. If possible, the third level (denoted by an arabic number)
lists all solutions divided into subclasses. Note that all variables not
further specified in the third level still have to satisfy the
conditions stated in the first and second level.

All of these solutions to the projector condition are valid for any
$\k>0$. By taking the continuous tensor product one has to choose a
different set of coefficients~$(\a,\ldots,\l)$, i.\,e., the coefficients
are promoted to functions of $\k$. Since the change of basis from
discrete to continuous oscillators involves integrations it seems to be
necessary to demand integrability of the coefficient functions. Later on,
we will demand additional restrictions for certain subclasses of projectors.
Apart from that, further restrictions would amount to specify the class
of allowed string fields~\cite{Fuchs:2002zz}. We will not impose more
restrictions now, but instead leave this subject open for further work.%
\footnote{In principle, one can even try to employ different types of
projectors for different $\k$'s.}

\noindent
{\bf Representation as states.} In order to elucidate which states
actually are parametrized by the projectors in the last paragraph, we
first translate the ansatz~(\ref{eq:projans}) into string fields. The
corresponding states will then be extracted, which, with the coefficients
as in the above classification, are projector states.

The inverse Moyal-Weyl map determines the string field (localized at one
fixed value of~$\k$)
\begin{equation}
\label{eq:projanssf}
\begin{split}
  P[\vec{x}(\k),\vec{y}(\k)] & = \a + \sfrac{(\b+\g)\reg}{2} +
    \sfrac{\l\reg^2}{4} + \sfrac{2\b+\l\reg}{4\th}x_e y_o -
    \sfrac{2\g+\l\reg}{4\th}x_o y_e \\
  & \quad{}+ \sfrac{\ic\de}{2\th}x_e x_o - \sfrac{\ic\ve}{2\th}y_e y_o
    + \sfrac{\ic\vp}{2\th}x_e y_e - \sfrac{\ic\eta}{2\th}x_o y_o
    - \sfrac{\l}{4\th^2}x_e y_o x_o y_e
\end{split}
\end{equation}
corresponding to the ansatz~(\ref{eq:projans}) if one makes use of the
fact that, e.\,g., $x_e\star y_o = x_e y_o + \th\reg$. Here, all star
products are already evaluated in terms of the ordinary Grassmann product.
The orthogonality of the sectors of the star algebra with different $\k$
allows us to restrict ourselves to one $\k$ for the time being.

In order to translate the string field~(\ref{eq:projanssf}) into a linear
combination of Fock space states, we introduce the generating state
\begin{equation}
  |G\> = \exp\Big[ -\frac{1}{2\reg}\sum_\k \big( \vec{\ps}^{+\dag}_\k
    \cdot \vec{\ps}^{-\dag}_\k + \sqrt{2}\ic\vec{\l}(\k)\cdot
    \vec{\ps}^{-\dag}_\k + \sqrt{2}\vec{\ps}^{+\dag}_\k\cdot\vec{\mu}(\k)
    + \ic\vec{\l}(\k)\cdot\vec{\mu}(\k) \big)\Big]\dv\, ,
\end{equation}
which can be contracted with (the discretized version of) eq.~(\ref{eq:Xev})
to give
\begin{equation}
  \<\vec{X}|G\> \propto \exp\Big(\frac{1}{2\reg}\sum_\k (\vec{\l}(\k)\cdot
    \vec{x}(\k) - \vec{y}(\k)\cdot\vec{\mu}(\k))\Big)
\end{equation}
up to a proportionality constant independent of $\reg, \k$. Neglecting this
constant, $x_o y_o$, e.\,g., for a single~$\k$ corresponds to $2\ic\reg\exp
\big(-\sfrac{1}{2\reg}(\ps^{+\dag}_{\k,e}\ps^{-\dag}_{\k,e}-\ps^{+\dag}_{\k,o}
\ps^{-\dag}_{\k,o})\big)\dv$. Thus, the state corresponding to~%
(\ref{eq:projanssf}) reads
\begin{equation}
\label{eq:projansst}
\begin{split}
  |P_\k\> & = \big(f(\a,\b,\g,\l) - \sfrac{\l\reg^2}{\th^2}
    - \sfrac{\vp\reg}{\th} + \sfrac{\eta\reg}{\th} \big)\dv
    - \sfrac{1}{2\reg}\big(f(\a,\b,\g,\l) + \sfrac{\l\reg^2}{\th^2}
    + \sfrac{\vp\reg}{\th} + \sfrac{\eta\reg}{\th}\big)
    \ps^{+\dag}_{\k,e}\ps^{-\dag}_{\k,e}\dv \\
  & \quad {}-\sfrac{1}{2\reg}\big(f(\a,\b,\g,\l) + \sfrac{\l\reg^2}{\th^2}
    - \sfrac{\vp\reg}{\th} - \sfrac{\eta\reg}{\th}\big)
    \ps^{+\dag}_{\k,o}\ps^{-\dag}_{\k,o}\dv + \sfrac{2\b+\l\reg}{2\th}\,
    \ic\,\ps^{+\dag}_{\k,o}\ps^{-\dag}_{\k,e}\dv - \sfrac{2\g+\l\reg}
    {2\th}\,\ic\,\ps^{+\dag}_{\k,e}\ps^{-\dag}_{\k,o}\dv \\
  & \quad {}-\sfrac{\ic\de}{\th}\ps^{-\dag}_{\k,e}\ps^{-\dag}_{\k,o}\dv
    - \sfrac{\ic\ve}{\th}\ps^{+\dag}_{\k,e}\ps^{+\dag}_{\k,o}\dv
    + \sfrac{1}{4\reg^2}\big(f(\a,\b,\g,\l)-\sfrac{\l\reg^2}{\th^2}
    + \sfrac{\vp\reg}{\th} - \sfrac{\eta\reg}{\th}\big)
    \ps^{+\dag}_{\k,e}\ps^{-\dag}_{\k,e}\ps^{+\dag}_{\k,o}
    \ps^{-\dag}_{\k,o}\dv
\end{split}
\end{equation}
with
\begin{equation}
  f(\a,\b,\g,\l) = \a+\sfrac{(\b+\g)\reg}{2}+\sfrac{\l\reg^2}{4}\, .
    \label{eq:fdef}
\end{equation}
It is remarkable that our classification admits projectors which are not
neutral and not even homogeneous in their $U(1)$ charge (those with $\de
\neq 0$ or $\ve\neq 0$).

\noindent
{\bf Reality condition.} Admissible string fields are subject to a reality
condition with respect to star conjugation, cf.~\cite{Witten:1985cc,%
Gaberdiel:1997ia}: The hermitean conjugate has to equal the BPZ conjugate
of a given state. In our case~(\ref{eq:projansst}) this leads to the condition
that $\de$, $\ve$, $\vp$, $\eta$, and $\l$ are real, as well as the linear
combination $f(\a,\b,\g,\l)$. Furthermore, the real parts of $\b$ and $\g$
are restricted to be
\begin{equation}
  \text{Re }\b = \text{Re }\g = -\sfrac{\l\reg}{2}\, .
\end{equation}
Imposing the reality condition in addition to eqs.~(\ref{eq:pc}) reduces
the number of solutions in all classes:
\begin{itemize}
  \item[(I'):] $2\a+\reg\b+\reg\g =1$
  \begin{itemize}
    \item[(a'):] $\de\ne 0$ $\vee$ $\ve\ne 0$ $\Rightarrow$ $\l=0\,$,
                       $\de\ve=\b\g+\vp\h=\sfrac{1}{\reg^2}(\a^2-\a)\,$,
                       $\text{Re }\b=\text{Re }\g=0\,$.
    \item[(b'):] empty.
  \end{itemize}
  \item[(II'):] $\b=\g$
  \begin{itemize}
    \item[(a'):] $\vp\ne 0$ $\vee$ $\h\ne 0$ $\Rightarrow$
              $\a=\sfrac{1}{2}-\reg\g\,$, $\de\ve=\g^2-\sfrac{1}
              {4\reg^2}\,$, $\vp\h=-\sfrac{1}{4\reg^2}\,$.
    \begin{itemize}
      \item[(a1'):] $\de\ve\ne 0$ $\Rightarrow$ $\l=0\,$, $\text{Re }\b=
                    \text{Re }\g=0\,$.
      \item[(a2'):] empty.
      \item[(a3'):] $\de=\ve=0$ $\Rightarrow$
        $(\a,\g,\l)\in\{(0,\sfrac{1}{2\reg},-\sfrac{1}{\reg^2}),
                        (1,-\sfrac{1}{2\reg},\sfrac{1}{\reg^2})\}\,$.
    \end{itemize}
    \item[(b'):] $\vp=\h=0$ $\Rightarrow$ $\g=\sfrac{1-\a}{\reg}$ $\vee$
        $\g=-\sfrac{\a}{\reg}\,$, $(\l=\sfrac{1-2\a}{\reg^2}-\sfrac{2\g}
          {\reg}\, ,\,\text{Re }\a=\sfrac{1}{2})$ $\vee$ $(\l=-\sfrac{2\g}
          {\reg}\, ,\,\g\in\R)$, \\
        $\de\ve=\sfrac{\a^2-\a}{\reg^2}\,$.
    \begin{itemize}
      \item[(b1'):] $\de\ve\ne 0\,$.
      \item[(b2'):] empty.
      \item[(b3'):] $\de=\ve=0\:\Rightarrow\:(\a,\g,\l)\in\{(0,\frac{1}
        {\reg},-\frac{2}{\reg^2}),(0,0,0),(1,0,0),(1,-\frac{1}{\reg},
        \frac{2}{\reg^2})\}\,$.
    \end{itemize}
  \end{itemize}
\end{itemize}
From now on, we will switch to bra states, which makes the identification
of surface states somewhat easier.

\section{Squeezed state projectors}\label{sec:sqstate}
\noindent
In this section, we restrict to the subclass of squeezed state projectors
which are BPZ-real and diagonal in the $\k$-basis. A general squeezed state
in the fermionic sector has the form
\begin{equation}
  \<S| = \dvb\exp\Big[ \frac{1}{2}\sum_{m,n}(\ps^+_m S_{mn}\ps^-_n +
    \ps^+_m P_{mn}\ps^+_n + \ps^-_m M_{mn}\ps^-_m) \Big] \, ,
  \label{eq:sqstdo}
\end{equation}
where we admit also projectors with indefinite $U(1)$ charge. We will see
below that such projectors are included in our classification. Furthermore,
we neglect possible normalization factors.

\noindent
{\bf Squeezed state projectors in the diagonal basis.} In the continuous
basis, the squeezed states feature two $\k$-integrations in the exponent.
Note, however, that we restricted our classification to projectors which
are already projectors for each $\k$. Exactly if this is the case, $S_{mn}$,
$(EM)_{mn}$, $(E^{-1}P)_{mn}$, and the Neumann coefficients $N^{rs}_{mn}$
are simultaneously diagonalizable;\footnote{Here, $E$ is the matrix with
components $E_{mn}=n\de_{mn}$ introduced in section~\ref{sec:moyalvertex}.}
and the exponent can be reduced to contain only one $\k$-integration. Namely,
for $S_{mn}$, the twist properties of $v^+_n(\k)$,
\begin{equation}
  v^+_{2n}(-\k) = -v^+_{2n}(\k)\, ,\qquad v^+_{2n+1}(-\k) = v^+_{2n+1}(\k)
    \, ,
\end{equation}
imply that the even and odd parts are also eigenvectors of $S_{mn}$,
\begin{align}
  \sum_n S^{\phantom{+}}_{2m,2n}v^+_{2n}(\k) & = S_{ee}(\k) v^+_{2m}(\k)
  \, , & \sum_n S^{\phantom{+}}_{2m+1,2n}v^+_{2n}(\k) & = S_{oe}(\k)
  v^+_{2m+1}(\k)\, , \notag \\
  \sum_n S^{\phantom{+}}_{2m,2n+1}v^+_{2n+1}(\k) & = S_{eo}(\k) v^+_{2m}
  (\k)\, , & \sum_n S^{\phantom{+}}_{2m+1,2n+1}v^+_{2n+1}(\k) & = S_{oo}
  (\k) v^+_{2m+1}(\k)\, .
\end{align}
Here, $S_{ee}$, $S_{oe}$, $S_{eo}$, and $S_{oo}$, denote the corresponding
eigenvalues.\footnote{Note that for consistency, $S_{ee}$ and $S_{oo}$
have to be even functions of~$\k$, and $S_{eo}$ and $S_{oe}$ have to be
odd functions of~$\k$.} Thus, eq.~(\ref{eq:orthcmpl}) guarantees that
\begin{equation}
  \sum_{m,n}v^-_{2m}(\k')S^{\phantom{+}}_{2m,2n}v^+_{2n}(\k) = \sfrac{1}
    {2} S_{ee}(\k)\de(\k-\k') \label{eq:vSv}
\end{equation}
and similar relations hold for the other components. If one rewrites
$\ps^+_m S_{mn}\ps^-_n$ in terms of the continuously moded operators,
the delta distributions on the right-hand side of eq.~(\ref{eq:vSv})
can be used to remove the $\k'$-integration. In the case of $P_{mn}$,
we have
\begin{subequations}
\begin{equation}
\begin{split}
  \sum_n\big(E^{-1}P\big)_{2m+1,2n} & v^{-\phantom{\dag}}_{2n}(\k) =
    p_{oe}(\k)v^-_{2m+1}(\k)\\
    \Longrightarrow\quad & \sum_{m,n}\big( v^-(\k')E
    \big)_{2m+1}\big(E^{-1}P\big)_{2m+1,2n}v^{-\phantom{\dag}}_{2n}(\k)
    = \sfrac{1}{2}p_{oe}(\k)\de(\k-\k')\, ,
\end{split}
\end{equation}
\begin{equation}
\begin{split}
  \sum_n\big(E^{-1}P\big)_{2m,2n+1} & v^{-\phantom{\dag}}_{2n+1}(\k) =
    p_{eo}(\k)v^-_{2m}(\k) \\
    \Longrightarrow\quad & \sum_{m,n}\big( v^-(\k')E
    \big)_{2m}\big(E^{-1}P\big)_{2m,2n+1} v^{-\phantom{\dag}}_{2n+1}(\k)
    = \sfrac{1}{2}p_{eo}(\k)\de(\k-\k')\, .
\end{split}
\end{equation}
\end{subequations}
Similar arguments hold for the eigenvalues $m_{eo}$ and $m_{oe}$ of
$(EM)_{mn}$. Then, we can rewrite the squeezed state as
\begin{equation}
  \< S| = \dvb\exp\Big[\frac{1}{2}\int_0^\infty d\k (\vec{\ps}^+_\k
    \cdot S(\k)\cdot\vec{\ps}^-_\k -\ic P_{eo}(\k)\ps^+_{e,\k}
    \ps^+_{o,\k} -\ic M_{eo}(\k)\ps^-_{e,\k}\ps^-_{o,\k})\Big]
    \, . \label{eq:sqstate}
\end{equation}
with $S(\k)=\left(\begin{smallmatrix} S_{ee} & -\ic S_{eo}\\ -\ic S_{oe} &
-S_{oo}\end{smallmatrix}\right)$, $P_{eo}(\k)=p_{oe}(\k)-p_{eo}(\k)$ and
$M_{eo}(\k)=m_{oe}(\k)-m_{eo}(\k)$.

\noindent
{\bf Squeezed state conditions.} There is a simple criterion of whether a
state~(\ref{eq:projansst}) is a squeezed state. This question is
particularly important when it comes to the interpretation of the states
in our classification as surface states. To begin with, it should be
remarked that a squeezed state includes a nonvanishing constant term from
the expansion of the exponential (which in the canonical normalization of
the state equals 1). As a consequence, a necessary condition 
for~(\ref{eq:projansst}) to be a squeezed state is
\begin{equation}
  g(\a,\b,\g,\vp,\eta,\l) := f(\a,\b,\g,\l) - \sfrac{\l\reg^2}{\th^2}
    - \sfrac{\vp\reg}{\th} + \sfrac{\eta\reg}{\th}\neq 0\, .
  \label{eq:sqnec}
\end{equation}
If this is the case, we may factorize the state into $g(\a,\b,\g,\vp,
\eta,\l)$ times a canonically normalized state. The condition for the
latter to take the exponential form~(\ref{eq:sqstate}) is that the
$\ps^{+\dag}_{\k,e}\ps^{-\dag}_{\k,e}\ps^{+\dag}_{\k,o}\ps^{-\dag}_{\k,o}
\dv$-term in~(\ref{eq:projansst}), normalized appropriately, agrees with (the
BPZ conjugate of) the term quartic in the $\ps$'s from the exponential,
i.\,e., with
\begin{equation}
  \sfrac{1}{4\reg^2}(-S_{ee} S_{oo} + S_{oe} S_{eo} + P_{eo} M_{eo})
    \ps^{+\dag}_{\k,e}\ps^{-\dag}_{\k,e}\ps^{+\dag}_{\k,o}
    \ps^{-\dag}_{\k,o}\dv\, ,
\end{equation}
where
\begin{subequations}
\label{eq:Seeetc}
\begin{gather}
  S_{ee} = \big(f(\a,\b,\g,\l) + \sfrac{\l\reg^2}{\th^2}
    + \sfrac{\vp\reg}{\th} + \sfrac{\eta\reg}{\th}\big)/g(\a,\b,\g,\vp,\eta,\l)
    \, , \\
  S_{oo} = - \big(f(\a,\b,\g,\l) + \sfrac{\l\reg^2}{\th^2}
    - \sfrac{\vp\reg}{\th} - \sfrac{\eta\reg}{\th}\big)/g(\a,\b,\g,\vp,\eta,\l)
    \, , \\
  S_{oe} = \big(2\b\reg+\l\reg^2\big)/(\th g(\a,\b,\g,\vp,\eta,\l))\, , \\
  S_{eo} = -\big(2\g\reg+\l\reg^2\big)/(\th g(\a,\b,\g,\vp,\eta,\l))\, , \\
  M_{eo} = 2\de\reg/(\th g(\a,\b,\g,\vp,\eta,\l))\, , \\
  P_{eo} = 2\ve\reg/(\th g(\a,\b,\g,\vp,\eta,\l))
\end{gather}
\end{subequations}
from~(\ref{eq:projansst}). This yields the necessary and sufficient
condition
\begin{equation}
  \a\l = \vp\eta + \b\g - \de\ve\, ; \label{eq:sqstcondgen}
\end{equation}
a state satisfying this condition (and with $g\neq 0$) is a squeezed state
of the form~(\ref{eq:sqstate}). For a projector state $|P\>_\k$ satisfying
eq.~(\ref{eq:pc8}) this condition simplifies to
\begin{equation}
  0 = \l(4f(\a,\b,\g,\l) - 1)\quad\Longrightarrow\quad \l=0\:\:\vee\:\:
    f(\a,\b,\g,\l)=\sfrac{1}{4}\, . \label{eq:sqstcond}
\end{equation}
It may serve as a simple check of~(\ref{eq:sqstate}) and~(\ref{eq:Seeetc})
that for $\a=1$, all other parameters zero, we obtain the identity state
\begin{equation}
  \<\Ical| = \dvb\exp\Big(\frac{1}{2}\int_0^\infty d\k\,\vec{\ps}^+_\k
    \cdot\vec{\ps}^-_\k\Big) \label{eq:idsqst}
\end{equation}
if we take the tensor product over all~$\k$.

Imposing conditions~(\ref{eq:sqnec}) and~(\ref{eq:sqstcond}) in addition
to the classification of real projectors in the last section reduces the
number of solutions in some cases:
\begin{itemize}
  \item[(I''):] $2\a+\reg\b+\reg\g =1$
  \begin{itemize}
    \item[(a''):] $\de\ne 0$ $\vee$ $\ve\ne 0$ $\Rightarrow$ $\l=0\,$,
                  $\de\ve=\b\g+\vp\h=\sfrac{1}{\reg^2}(\a^2-\a)\,$,
                  $\vp\neq\eta+\sfrac{\th}{2\reg}\,$, $\text{Re }\b=
                  \text{Re }\g=0$.
    \item[(b''):] empty.
  \end{itemize}
  \item[(II''):] $\b=\g$
  \begin{itemize}
    \item[(a''):] $\vp\ne 0$ $\vee$ $\h\ne 0$ $\Rightarrow$
                  $\a=\sfrac{1}{2}-\reg\g\,$, $\de\ve=\g^2-\sfrac{1}
                  {4\reg^2}\,$, $\vp\h=-\sfrac{1}{4\reg^2}\,$.
    \begin{itemize}
      \item[(a1''):] $\de\ve\ne 0$ $\Rightarrow$ $\l=0\,$, $\vp\neq\eta+
        \sfrac{\th}{2\reg}\,$, $\text{Re }\b=\text{Re }\g=0$.
      \item[(a2''):] empty.
      \item[(a3''):] $\a=0\, ,\,\g=\sfrac{1}
        {2\reg}\, ,\,\l=-\sfrac{1}{\reg^2}\, ,\,\vp\neq\eta+\sfrac{\th}
        {4\reg} + \sfrac{1}{\th\reg}\,$.
    \end{itemize}
    \item[(b''):] $\vp=\h=0$
    \begin{itemize}
      \item[(b1''):] $\de\ve\ne 0\,$ $\Rightarrow$ $(\a,\g,\l)\in
        \{(\sfrac{1}{2}, -\sfrac{1}{2\reg}, \sfrac{1}{\reg^2}),
        (-\sfrac{1}{2}, \sfrac{3}{2\reg}, -\sfrac{3}{\reg^2})\}\,$.
      \item[(b2''):] empty.
      \item[(b3''):] $\de=\ve=0\:\Rightarrow\:\a=1\, ,\,\g=0\, ,\,\l=0\,$.
    \end{itemize}
  \end{itemize}
\end{itemize}
It is remarkable that there is only one class (namely~(IIa3'')) of real
neutral (i.\,e., with $\de=\ve=0$) squeezed state projectors apart from
the identity~(\ref{eq:idsqst}) in~(IIb3''). It will be demonstrated in the next
paragraph that this class contains the generalized butterfly states. We
will enlarge on this one-parameter family in the next section.

\noindent
{\bf Generalized butterfly states as Moyal projectors.} We will now show
that the best-known class of surface state projectors, the generalized
butterfly states, are contained in the above classification. This family~%
\cite{Gaiotto:2002kf} is parametrized by a parameter $a\in[0,2]$, and
in the continuous basis all members are proportional to~\cite{Fuchs:2002zz}
\begin{equation}\label{eq:gbutt}
  \< B_a|=\dvb\exp\Big(-\frac{1}{2}\int_0^\infty d\k\,\vec{\ps}^+_\k
    \begin{pmatrix}
      \sfrac{2\tanh\big(\sfrac{\pi\k(2-a)}{4a}\big)-\th(\k)}{2\tanh\big(
        \sfrac{\pi\k(2-a)}{4a}\big)+\th(\k)} & 0\\
      0 & \sfrac{2\coth\big(\sfrac{\pi\k(2-a)}{4a}\big)-\th(\k)}{2\coth
        \big( \sfrac{\pi\k(2-a)}{4a}\big)+\th(\k)}
    \end{pmatrix}\vec{\ps}^-_\k\Big)\, .
\end{equation}
The sliver is recovered in the limit $a\rightarrow 0$; the canonical
butterfly state is obtained for $a=1$, and the limit $a\to 2$ gives the
so-called nothing state. A straightforward computation (to be described
below) yields that this family of states is given by $\a=0$, $\b=\g=
\frac{1}{2\reg}$, $\de=\ve=0$, $\eta=\frac{1}{2 h_a \reg}$, $\vp=
-\frac{h_a}{2\reg}$, and $\l=-\frac{1}{\reg^2}$ with
\begin{equation}
  h_a(\k) = \tanh\big(\sfrac{\pi\k(2-a)}{4a}\big)
\end{equation}
in our classification scheme. Hence, these projectors belong to the
subclass~(IIa3'').

\noindent
{\bf Relation to bosonic projectors.} It is interesting to note that there is 
a one-to-one correspondence between (neutral) squeezed state projectors of 
our fermionic first order system and those of the bosonic CFT in momentum 
basis. Namely, an inspection of the corresponding squeezed state formulas 
yields that a fermionic projector with coefficient matrix $S(\k)$ leads to a 
bosonic projector with the same coefficient matrix (and vice versa).\footnote{%
In the discrete basis, a fermionic squeezed state projector with coefficients 
$S_{mn}^{\text{ferm.}}$ maps to a bosonic squeezed state projector with 
coefficients $S_{mn}^{\text{bos.}}=\frac{1}{2}\sqrt{\frac{n}{m}}
S_{mn}^{\text{ferm.}}$.} In the light of this observation, it is not
surprising that all our (neutral) squeezed state projectors, i.e., those
from (IIa3''), identically fulfill the projector condition of~%
\cite{Fuchs:2002zz}, eq.~(3.55). Thus, the analysis in other representations
(such as the half-string or discrete Moyal representations) given in this
reference carry over to our fermionic first order system. There is some hope
that the above correspondence can be generalized to the case of non-squeezed 
state projectors.

\noindent
{\bf Inverse transformation to operators.} Sometimes it is useful to describe
a given squeezed state in terms of operators under the Moyal-Weyl map, e.\,g.,
for the computation of star products or for testing whether it is contained
in our above classification (and thus a projector). For simplicity, we
restrict to uncharged states (with $M_{eo}=P_{eo}=0$); however, the
derivation can be trivially expanded to the general case. After giving the
appropriate formulas, we will describe this procedure for the example of the
sliver state.

Instead of solving eq.~(\ref{eq:Seeetc}) for $\a,\ldots,\l$, it is much
more convenient to transform the squeezed state projector (of the
form~(\ref{eq:sqstate})) under consideration to a string functional. This in 
general leads to a string field of the form
\begin{equation}
  \Psi_S[\vec{x}(\k), \vec{y}(\k)] = \big(\prod_\k f(\a,\b,\g,\l)\big)
    \exp\big(\sfrac{\ic}{2}\int d\k\,\vec{x}(\k)\cdot T(\k)\cdot\vec{y}
    (\k)\big) \, , \label{eq:sqstsf}
\end{equation}
where $T(\k)=\left(\begin{smallmatrix}T_{ee} & T_{eo}\\ T_{oe} & T_{ee}
\end{smallmatrix}\right)= \left(\frac{S(\k)-{\mathbbm 1}}{S(\k)+
{\mathbbm 1}}\right)^t$, and the proportionality factor was fixed by
comparison with eq.~(\ref{eq:projanssf}). Then, expanding the exponential
and comparing the coefficients with~(\ref{eq:projanssf}) yields
\begin{subequations}\label{eq:bcktrpar}
\begin{align}
  \b & = \frac{f(\a,\b,\g,\l)\th(\k)}{\reg}\big( \sfrac{1}{4}\det T +
           \ic T_{eo}\big) \, , \\
  \g & = \frac{f(\a,\b,\g,\l)\th(\k)}{\reg}\big( \sfrac{1}{4}\det T -
           \ic T_{oe}\big) \, , \\
  \vp & = \frac{f(\a,\b,\g,\l)\th(\k)}{\reg}\, T_{ee}\, , \\
  \eta & = -\frac{f(\a,\b,\g,\l)\th(\k)}{\reg}\, T_{oo}\, , \\
  \l & = -\frac{f(\a,\b,\g,\l)\th(\k)^2}{\reg^2}\det T\, .
\end{align}
 These five equations for the four components of~$T$ are consistent only
if the squeezed state condition~(\ref{eq:sqstcondgen}) holds. From the
definition~(\ref{eq:fdef}), we finally read off
\begin{equation}
  \a = f(\a,\b,\g,\l)\big( 1 + \sfrac{\ic\th(\k)}{2}(T_{oe}-T_{eo}) -
    \sfrac{\th^2}{4}\det T\big) \, .
\end{equation}
\end{subequations}
For squeezed state projectors, the overall scale $f(\a,\b,\g,\l)$ is fixed
to be $\frac{1}{4}$ if $\det T\neq 0$. In all other cases, it has to be
determined from the exact normalization of the string field~(\ref{eq:sqstsf}).

Let us now demonstrate the above procedure in the case of the sliver,
i.\,e., the generalized butterfly state with $a\to 0$. Since in this
limit, $h_a\to 1$, we obtain from~(\ref{eq:gbutt}) the ket state
(w.\,r.\,t.\ the canonically normalized star product)
\begin{equation}\label{eq:sliver2}
  |\Xi\> = \exp\big[ -\de(0)\int\!d\k\,\ln (\frac{4\th^2(\k)}
    {(\th(\k)+2)^2})\big] \exp\Big[ -\frac{1}{2}\int_0^\infty d\k\,
    \frac{\th(\k)-2}{\th(\k)+2}\,\vec{\psi}_\k^{+\dag}\cdot
    \vec{\psi}_\k^{-\dag}\Big]\dv\, .
\end{equation}
If we contract this with the position eigenstate~(\ref{eq:Xev}) the
string field functional for the sliver turns out to be
\begin{equation}
  \Psi_{\Xi}(\vec{x}(\k),\vec{y}(\k)) = \<\vec{X}|\Xi\>
    \propto\exp\Big[-\ic\int_0^\infty d\k\,\frac{\vec{x}(\k)\cdot\vec{y}
    (\k)}{\th(\k)}\Big] \, .
\end{equation}
This leads to the values $\a=0$, $\b=\g=\frac{2f}{\reg}$, $\vp=-\frac{2f}
{\reg}$, $\eta=\frac{2f}{\reg}$ as well as $\l=-\frac{4f}{\reg^2}$. The
squeezed state condition requires~$f=\frac{1}{4}$ in accordance with the
discussion on generalized butterflies (the proportionality factor is an
infinite product as in~(\ref{eq:sqstsf})).

\section{Identification of surface states}\label{sec:surfst}
\noindent
Some of the projectors of our classification scheme described in the preceding 
section can be identified with well-known surface states. Even though this
identification is rather technical in general, it is possible and will
be done for some cases in this section.

To begin with, we note that surface states have a representation as
exponentials of (linear combinations of) Virasoro generators, i.\,e.,
we can limit our search for surface states to squeezed states of the
form~(\ref{eq:sqstate}). However, not all such states are eligible,
namely states with $P_{mn}\neq 0$ or $M_{mn}\neq 0$ cannot be expressed
in terms of (neutral) Virasoro generators. This restricts our search to
the one-parameter class given by
\begin{equation}
  \a=0\, , \quad \b=\g=\frac{1}{2\reg}\, , \quad \de=\ve=0\, , \quad
    \eta=\frac{1}{2r\reg}\, , \quad \vp=-\frac{r}{2\reg}\, , \quad
    \mbox{and}\quad\l=-\frac{1}{\reg^2}\, , \label{eq:ssparams}
\end{equation}
where $r(\k)$ is a free odd function of~$\k$ (the latter requirement
ensures that $S_{ee}$ and $S_{oo}$ are even functions). All projectors
in this class have diagonal matrices~$S$, i.\,e., $S_{eo}=S_{oe}=0$.
This class also contains the generalized butterfly states for the special
choice $r=h_a$, cf.\ section~\ref{sec:projec}.

\noindent
{\bf General method.} A surface state $\<\S^f|$ is determined by a map
$f:H\to\S$ from the canonical upper half disk $\{|z|\leq 1\}\subset\C$
onto a Riemann surface~$\S$ with boundary and the requirement that
\begin{equation}
  \<\S^f|\phi\> = \< f\circ\phi(0)\>_\S \label{eq:surfstdef}
\end{equation}
for all Fock space states~$|\phi\>$ of weight~$h$ in the boundary CFT
with corresponding operators $\phi(z)$. The correlation function
$\<\,\>_\S$ is evaluated on~$\S$, and $f\circ\phi(0)=\big(f'(0)\big)^h
\phi(f(0))$ is the conformal transform of~$\phi$ by the map~$f$.

Now, let us evaluate the correlation function~(\ref{eq:surfstdef}) for
a surface state of the form~(\ref{eq:sqstdo}) with~$M_{mn}=P_{mn}=0$
and for~$|\phi\>=\ps^+(z)\ps^-(w)\ps^+(0)\dv$,\footnote{The $U(1)$ charges
of the insertions of the correlation functions have to sum up to $+1$ in
order to give a nonvanishing result. -- The somewhat unusual factors of~2
in the next two formulas stem from our normalization~(\ref{eq:comrel}).}
\begin{equation}
  \dvb\exp\Big[\sfrac{1}{2}\sum_{m,n>0}\ps^+_m S^{\phantom{+}}_{mn}
    \ps^-_n\Big] \sum_{k=-\infty}^\infty \frac{\ps^+_k}{z^k}\sum_{l
    = -\infty}^{-1} \frac{\ps^-_l}{w^{l+1}}\dv = 2\sum_{m,n>0} S_{mn}
    z^n w^{m-1} + \frac{2}{z-w}\frac{z}{w}\, .
\end{equation}
On the other hand, according to eq.~(\ref{eq:surfstdef}), this correlation
function should be equal to\footnote{Recall that the second
factor originates from the nontrivial background charge of the system~%
\cite{Kling:2003sb}.}
\begin{equation}
  \< f\circ\phi\>_\S = \frac{2 f'(w)}{f(z)-f(w)}\frac{f(z)-f(0)}
    {f(w)-f(0)}\, ;
\end{equation}
and we can use $SL(2,\R)$ invariance to fix $f(0)=0$, $f'(0)=1$, and
$f''(0)=0$. Then, (one half of) the nonsingular part of the correlator
becomes~\cite{Okuda:2002fj,Bars:2003gu} 
\begin{equation}
  S(z,w) := \sum_{m,n>0} S_{mn} z^n w^{m-1} = \frac{f'(w)}{f(z)-f(w)}
    \frac{f(z)}{f(w)} - \frac{1}{z-w}\frac{z}{w}\, , \label{eq:Szw}
\end{equation}
therefore, deriving w.\,r.\,t.\, $z$ and choosing $w=0$,
\begin{equation}
  \frac{\pa}{\pa z}S(z,0) = -\frac{f'(z)}{f(z)^2} + \frac{1}{z^2}\, .
\end{equation}
This equation can be integrated to give
\begin{equation}
  f(z) = \frac{z}{z S(z,0) + 1} \label{eq:fS}
\end{equation}
for a candidate function~$f$ for the map defining the surface state~$\< S^f|$.
Here, the integration constant was chosen in such a way that for
$\< S^f|=\dvb$, we obtain the identity map $f(z)=z$.

Obviously, the defining map~$f(z)$ is encoded in the coefficients~$S_{mn}$
via~(\ref{eq:Szw}); we can extract a candidate function with the help of
eq.~(\ref{eq:fS}). From our classification of the projectors in the last
section, we know the $S_{mn}$ in terms of the eigenvalues $S_{ee}$,
$S_{oe}$, $S_{eo}$, and $S_{oo}$. Since all squeezed state projectors
under consideration have $S_{eo}=S_{oe}=0$, in fact only $S_{ee}$ and
$S_{oo}$ contribute. For the choice of parameters given in~%
(\ref{eq:ssparams}) these eigenvalues are parametrized by
\begin{equation}
  S_{ee}(\k) = -1 + \frac{\th(\k)}{\th(\k) + 2r(\k)}\, , \qquad
  S_{oo}(\k) = -1 + \frac{4}{2+r(\k)\th(\k)}\, . \label{eq:SeeSooPar}
\end{equation}
Now, $S(z,w)$ can be reconstructed
from these eigenvalues using~(\ref{eq:orthcmpl}), (\ref{eq:eogenf}), and~%
(\ref{eq:vSv}),
\begin{equation}
  S(z,w) = \frac{2}{w}\int_0^\infty d\k\Big[ f_{v_e^+}(\k,w)f_{v_e^-}
   (\k,z) S_{ee}(\k) + f_{v_o^+}(\k,w)f_{v_o^-}(\k,z)S_{oo}(\k)\Big]\, .
  \label{eq:Szw2}
\end{equation}
It should be noted that $f(z)$ is purely given by the $S_{oo}$-part, see~%
(\ref{eq:fS}) and
\begin{equation}
  S(z,0) = \int_0^\infty d\k\,\frac{\sinh \k Z}{\sinh\frac{\pi\k}
    {2}}S_{oo}(\k)\, . \label{eq:Sz0int}
\end{equation}
As in section~\ref{sec:moyalvertex}, $Z$ abbreviates~$\tan^{-1} z$. Given
a function $r(\k)$ such that the integral~(\ref{eq:Szw2}) converges, we
can determine a map~$f(z)$ via~(\ref{eq:fS}). Obviously, $f$ is an {\em
odd} function mapping the upper unit half-disk into some region of the
upper half-plane.

\noindent
{\bf Consistency conditions for surface states.} We will now dicuss two
consistency conditions on a map $f(z)$ obtained in this way:

If the map~$f(z)$ defines a surface state projector, eqs.~(\ref{eq:Szw})
and~(\ref{eq:Szw2}) necessarily have to agree (in fact, this condition
is also sufficient~\cite{Fuchs:2002zz}). Since $f(z)$ is constructed
solely from the $S_{oo}$-part in~(\ref{eq:Szw2}) and $S(z,w)$ as given
by~(\ref{eq:Szw2}) also contains~$S_{ee}$, this is a restriction on the
function~$r$ in~(\ref{eq:SeeSooPar}). However, we found it difficult
to solve this constraint for $r(\k)$ (e.\,g., by variational methods)
due to the nonlinearity of the right-hand side of the CFT expression~%
(\ref{eq:Szw}) in~$S_{oo}$.

Instead, it is possible to derive a second (necessary) consistency condition
for $f(z)$ which has to hold for any surface state projector diagonal in
the $\k$-basis and which can be used to determine the form of possible~$f$'s.
The starting point is the observation that the $S_{oo}$-part of eq.~%
(\ref{eq:Szw2}) is given by $S(z,0)$ and the $S_{ee}$-part can be
determined via (an integral of) $\frac{\pa}{\pa w}S(z,w)|_{w=0}$.
Thus, the full integral~(\ref{eq:Szw2}) can be computed via~$f(z)$.
Together with~(\ref{eq:Szw}), this gives a restriction on~$f(z)$.

Firstly, note that~(\ref{eq:Szw2}) can be rewritten as
\begin{equation}
  S(z,w) = \frac{1}{1+w^2}\Big[-S_1(W) + \frac{1}{2}S_1(Z+W) -
    \frac{1}{2}S_1(Z-W) + \frac{1}{2}S_2(Z+W) + \frac{1}{2}S_2(Z-W)
    \Big]\, , \label{eq:Szwsplit}
\end{equation}
again with $W=\tan^{-1} w$, $Z=\tan^{-1} z$, and
\begin{equation}
  S_1(Z) = \int_0^\infty d\k\,\frac{\sinh \k Z}{\sinh\sfrac{\pi\k}{2}}\,
    S_{ee}(\k)\, ,\qquad
  S_2(Z) = \int_0^\infty d\k\,\frac{\sinh \k Z}{\sinh\sfrac{\pi\k}{2}}\,
    S_{oo}(\k)\, .
\end{equation}
Obviously, $S_2(\tan^{-1} z)=\frac{1}{f(z)}-\frac{1}{z}$.

Next, it is easy to see that due to the form of~(\ref{eq:Szw2}), only
the $S_{ee}$-part will contribute to $\tilde{S}(Z):=\frac{\pa}{\pa w}
(1+w^2)S(z,w)|_{w=0} = \frac{\pa}{\pa w}S(z,w)|_{w=0}$; in addition, one can
recover $S_1(Z)$ fully from $\int_0^Z dZ' (\tilde{S}(Z') + \tilde{S}_0)$
with some additive constant~$\tilde{S}_0$. Eq.~(\ref{eq:Szwsplit}) ensures
that the contributions of this constant to $S(z,w)$ cancel.

Following this prescription, we have to compute 
$\frac{\pa}{\pa w}S(z,w)|_{w=0}$ from~%
(\ref{eq:Szw}). A direct computation with our choice $f(0)=0$, $f'(0)=1$,
$f''(0)=0$ leads to a difference of singularities. In order to avoid this
one can compute $\frac{\pa^2}{\pa z\pa w}S(z,w)$, set $w=0$, and integrate
with respect to~$z$. If one absorbs the integration constant into
$\tilde{S}_0$, the result is
\begin{equation}
  \frac{\pa}{\pa w}S(z,w)|_{w=0} = \frac{1}{f(z)^2} - \frac{1}{z^2}
    \quad\Longrightarrow\quad \tilde{S}(Z) = \frac{1}{F(Z)^2}-\frac{1}
    {\tan^2 Z}
\end{equation}
with $F(Z):=f(\tan Z)=f(z)$. Putting everything together, we obtain
from~(\ref{eq:Szwsplit}):
\begin{equation}
\begin{split}
  (1+w^2)S(z,w) = & -\int_0^W dZ'\,\tilde{S}(Z') + \frac{1}{2}\int_0^{Z+W}
    dZ'\,\tilde{S}(Z') - \frac{1}{2}\int_0^{Z-W}dZ'\,\tilde{S}(Z') \\
  & {}+\frac{1}{2}\Big(\frac{1}{F(Z+W)}-\frac{1}{\tan(Z+W)}\Big)
    +\frac{1}{2}\Big(\frac{1}{F(Z-W)}-\frac{1}{\tan(Z-W)}\Big) \, .
\end{split}
\end{equation}
This expression should be compared with eq.~(\ref{eq:Szw}). Finally,
a somewhat messy calculation leads to the following result:
\begin{equation}
  \frac{1}{2}\frac{1-F'(Z+W)}{F(Z+W)^2} + \frac{1}{2}\frac{1+F'(Z-W)}
    {F(Z-W)^2} = \frac{F''(W)}{F(W)} + \frac{1-F'(W)^2}{F(W)^2} +
    \frac{F''(W)}{F(Z)-F(W)} + \frac{F'(W)^2}{(F(Z)-F(W))^2}\, .
  \label{eq:surstcond}
\end{equation}
This condition restricts the allowed maps~$f(z)$. A short computation
shows that it is indeed satisfied for the generalized butterfly states
with $F(Z)=\frac{1}{a}\sin aZ$ as well as the identity state with $F(Z)
=\sfrac{1}{2}\tan 2Z$.\footnote{In fact, this relation also holds for
all wedge states with $F(Z)=\sfrac{1}{a}\tan aZ$ which can even be
obtained by an integration as in~(\ref{eq:fS}) and~(\ref{eq:Sz0int}).
However, the necessary and sufficient condition that eqs.~(\ref{eq:Szw})
and~(\ref{eq:Szw2}) have to agree is only satisfied for the identity
and butterfly states.} By a differentiation w.\,r.\,t.\ $Z$
and an integration over~$W$, eq.~(\ref{eq:surstcond}) can be transformed
into
\begin{equation}
  -\frac{F'(Z)F'(W)}{(F(Z)-F(W))^2} = \frac{1-F'(Z+W)}{2F(Z+W)^2} -
    \frac{1+F'(Z-W)}{2F(Z-W)^2}\, . \label{eq:surstcond2}
\end{equation}
The integration constant is fixed by the initial conditions $F(0)=0$,
$F'(0)=1$, and $F''(0)=0$. In this formulation, the left-hand side of
eq.~(\ref{eq:surstcond}) is a bosonic two-point function. It is easy
to see that condition~(\ref{eq:surstcond2}) is a refined version of
the surface state condition~(3.40) in~\cite{Fuchs:2002zz}, cf.\ also
eq.~(3.30) in this reference.

\noindent
{\bf Generalized butterfly states.} Exemplarily, we will demonstrate the
method in the case of the butterfly states~(\ref{eq:gbutt}) which are known
to be surface states. The integral we have to compute is
\begin{equation}
  S(z,0) = \int_0^\infty d\k\, \frac{\sinh \k Z}{\sinh\frac{\pi\k}{2}}\,
    \frac{2\coth\big(\frac{\pi\k(2-a)}{4a}\big)-\th(\k)}{2\coth\big(
    \frac{\pi\k(2-a)}{4a}\big)+\th(\k)}\, .
\end{equation}
Decomposing $\coth\big(\frac{\pi\k(2-a)}{4a}\big) = \frac{2-\tanh\big(
\frac{\pi\k}{2a}\big)\th(\k)}{2\tanh\big(\frac{\pi\k}{2a}\big)-\th(\k)}$,
we obtain
\begin{equation}
  S(z,0) = \int_0^\infty d\k\, \frac{\sinh \k Z}{\sinh\frac{\pi\k}{2}}
    \Big(1-\frac{\th(\k)-2\tanh\big(\frac{\pi\k}{2a}\big)}{\frac{\th(\k)}
    {2}-\frac{2}{\th(\k)}}\Big)\, .
\end{equation}
The first part can immediately be integrated with the help of
\begin{equation}
  \int_0^\infty dx\,\frac{\sinh ax}{\sinh bx} = \frac{\pi}{2b}\tan
    \frac{\pi a}{2b}\qquad\text{for}\qquad |\text{Re }a|<\text{Re }b\, .
\end{equation}
Using $\frac{\th(\k)}{2}-\frac{2}{\th(\k)}=-2\frac{1}{\sinh\big(\frac{\pi\k}{2}
\big)}$, we thus have to evaluate
\begin{equation}
\begin{split}
  S(z,0) & = z + \int_0^\infty d\k\,\sinh(\k Z) \big(\tanh\big(\sfrac{\pi\k}{4}
    \big)-\tanh\big(\sfrac{\pi\k}{2a}\big)\big) \\
  & = z + \int_{1/a}^{1/2}dy\int_0^\infty d\k\,\frac{\sinh(\k Z)}{\cosh\big(
    \frac{\pi\k y}{2}\big)^2}\frac{\pi\k}{2}\, ,
\end{split}
\end{equation}
where we have rewritten the difference of the two $\tanh$'s as an integral
over the derivative of the $\tanh$. In order to get rid of the simple
power of~$\k$ in the integrand, we rewrite the latter as a derivative, so
that we obtain the Gradshteyn/Ryzhik-integrable expression~\cite{GR1}
\begin{equation}
\begin{split}
  S(z,0) & = z + \frac{\pi}{2}\frac{\pa}{\pa Z}\int_{1/a}^{1/2}dy
    \int_0^\infty d\k\,\frac{\cosh(\k Z)}{\cosh\big( \frac{\pi\k y}{2}
    \big)^2} \\
  & = z + \frac{\pa}{\pa Z}\int_{1/a}^{1/2} dy\,\frac{\frac{Z}{y^2}}
    {\sin\frac{Z}{y}}\, .
\end{split}
\end{equation}
But since for a function $F\big(\sfrac{Z}{y}\big)$, $\frac{1}{y}\frac{\pa}
{\pa Z} F\big(\sfrac{Z}{y}\big) = -\frac{1}{Z} \frac{\pa}{\pa y}
F\big(\sfrac{Z}{y}\big)$, the $y$-integration is trivial to do,
and $S(z,0)$ simplifies to
\begin{equation}
  S(z,0) = z + \frac{a}{\sin aZ}-\frac{2}{\sin 2Z}
         = -\frac{1}{z} + \frac{a}{\sin aZ}\, .
\end{equation}
This yields the well-known expression
\begin{equation}
  f(z) = \frac{1}{a}\sin(a\tan^{-1} z)\, 
\end{equation}
for the map defining the generalized butterfly states.

\section{Conclusions}\label{sec:concl}
\noindent
In this paper we have considered star algebra projectors for a fermionic
first order system of weights $(1,0)$. To this end we have formulated the 
interaction vertex in the continuous Moyal basis. In this basis the 
oscillators form two Clifford algebras labeled by $\k\in\mathbbm{R}$. 
We have classified all projectors which factorize into projectors for
each value of $\k$ by employing a fermionic version of the Moyal-Weyl
map known from noncommutative field theories. BPZ-real squeezed state 
projectors which are neutral w.\,r.\,t.\ the $U(1)$ (ghost number) current 
turn out to be naturally parametrized by a single odd and integrable 
function of the parameter $\k$. We have shown that this class of
projectors contains the generalized butterfly states as a subclass. A 
method how to determine surface states in this class of projectors is 
given and a condition on the maps defining the shape of the surface 
state is derived. 

At least for the case of squeezed state projectors described above one 
can recover the full set of bosonic projectors. It is tempting to believe 
that this extends to non-squeezed state projectors which are neutral 
w.\,r.\,t.\ the $U(1)$ (ghost) current as well. 

It would be interesting to solve the consistency condition on the maps
defining surface states. Those solutions which indeed can be obtained from
integrations as in (5.9) and (5.10) will correspond to surface state
projectors which are diagonal in the $\k$-basis. It would be fascinating 
to see whether the generalized butterfly states and the identity state
are the only solutions, and if not, whether it is possible to obtain
surface state projectors without the singular property that the boundary
of the surface touches the midpoint. This is indeed the case for the
identity, but no finite rank projector with this property is known.
Moreover, it would be worthwhile to study the projectors we have
classified in a continuous half string basis along the lines of~%
\cite{Fuchs:2003wu}.

Another remarkable result we have obtained is that the normalization 
factors from rewriting the star product in the Moyal basis 
cancel between bosons and fermions in any even dimension. This is a 
consequence of the close relation of the corresponding Neumann
coefficients. Hence, up to an overall normalization factor the star
product of N=2 string field theory is a canonically normalized
continuous tensor product of Moyal-Weyl products. This is a
distinguished feature of this type of string field theory.

\subsubsection*{Acknowledgements}
\noindent
SU would like to thank Martin Schnabl and Yuji Okawa for discussions.
This work was done within the framework of the DFG priority program
``String Theory'' (SPP 1096). The work of SU was supported by a fellowship 
within the postdoc program of the German Academic Exchange Service (DAAD).

\end{document}